\documentclass[Afour,sageh,times]{sagej}
\usepackage[utf8]{inputenc}
\usepackage[usenames,dvipsnames,table]{xcolor}

\usepackage{moreverb,url,booktabs}

\usepackage[colorlinks,bookmarksopen,bookmarksnumbered,citecolor=red,urlcolor=red]{hyperref}

\definecolor{mygreen}{rgb}{0,0.6,0}
\definecolor{mygray}{rgb}{0.95,0.95,0.95}
\definecolor{mymauve}{rgb}{0.58,0,0.82}
\definecolor{mycodebg}{rgb}{1,0.96,0.83}
\definecolor{mytermbg}{rgb}{0.79,0.99,0.96}
\definecolor{mypythonbg}{rgb}{0.95,0.95,0.97}
\definecolor{myoutputbg}{rgb}{1.0,0.6, 0}
\definecolor{myexercise}{rgb}{0.52,0.80,0.98}
\definecolor{mymatlabbg}{rgb}{0.85,0.88,0.96}

\usepackage{algorithm}
\usepackage{algorithmic}
\usepackage{boxedminipage}

\usepackage{amsmath}
\usepackage{amsfonts}
\usepackage{caption}
\usepackage{subcaption}
\usepackage{multirow}

\newcommand{\TM}{\textsuperscript{\texttrademark}}

\usepackage{tikz}
\usepackage{pgfplots}
\usepackage{pgfmath}
\pgfplotsset{compat=1.17}

\definecolor{RYB1}{RGB}{166,206,227}  
\definecolor{RYB2}{RGB}{31,120,180}   
\definecolor{RYB3}{RGB}{178,223,138}  
\definecolor{RYB4}{RGB}{51,160,44}    
\definecolor{RYB5}{RGB}{251,154,153}  
\definecolor{RYB6}{RGB}{227,26,28}    
\definecolor{RYB7}{RGB}{253,191,111}  
\definecolor{RYB8}{RGB}{255,127,0}    
\definecolor{RYB9}{RGB}{202,178,214}  
\definecolor{RYB10}{RGB}{160,60,140}  
\definecolor{RYB11}{RGB}{255,255,153} 
\definecolor{RYB12}{RGB}{177,89,40}   

\definecolor{code}{rgb}{0, 0, 0}

\pgfplotscreateplotcyclelist{will}{%
black,every mark/.append style={fill=white},mark=*\\%
RYB4!50!black,every mark/.append style={fill=RYB4},mark=*\\%
RYB6!50!black,every mark/.append style={fill=RYB6},mark=*\\%
RYB8!50!black,every mark/.append style={fill=RYB8},mark=*\\%
RYB10!50!black,every mark/.append style={fill=RYB10},mark=*\\%
RYB12!50!black,every mark/.append style={fill=RYB12},mark=*\\%
black,every mark/.append style={fill=white!50!black},mark=*\\%
RYB2!50!black,every mark/.append style={fill=RYB2},mark=*\\%
RYB4!50!black,densely dashed,every mark/.append style={fill=RYB4},mark=*\\%
RYB6!50!black,densely dashed,every mark/.append style={fill=RYB6},mark=*\\%
RYB8!50!black,densely dashed,every mark/.append style={fill=RYB8},mark=*\\%
RYB10!50!black,densely dashed,every mark/.append style={fill=RYB10},mark=*\\%
RYB12!50!black,densely dashed,every mark/.append style={fill=RYB12},mark=*\\%
black,densely dashed,every mark/.append style={fill=white!50!black},mark=*\\%
RYB2!50!black,densely dashed,every mark/.append style={fill=RYB2},mark=*\\%
}

\usepackage{graphicx}

\usepackage{tikzit}

\tikzstyle{halo node}=[fill=red, draw=black, shape=circle, inner sep=-2]
\tikzstyle{interior node}=[fill=blue, draw=black, shape=circle, inner sep=-2]
\tikzstyle{kernel launch}=[fill=black, draw=black, shape=circle, inner sep=-2]
\tikzstyle{kernel entry}=[fill=none, draw=none, shape=circle, inner sep=0]
\tikzstyle{halo element}=[fill=red, draw=black, shape=rectangle]
\tikzstyle{interior element}=[fill=blue, draw=black, shape=rectangle]

\tikzstyle{interior element}=[-, fill=blue, draw=black, fill opacity=0.2, thick]
\tikzstyle{halo element}=[-, fill=red, draw=black, fill opacity=0.2, thick]
\tikzstyle{arrow}=[->, thick]
\tikzstyle{box outline}=[-, thick, rounded corners, fill=white]
\tikzstyle{dashed}=[-, draw={rgb,255: red,177; green,177; blue,177}, thick]

\begin{document}

\runninghead{Chalmers, Mishra, McDougall, and Warburton}

\title{HipBone: A performance-portable GPU-accelerated C++ version of the NekBone benchmark.}

\author{Noel Chalmers\affilnum{1}, Abhishek Mishra\affilnum{2}, Damon McDougall\affilnum{1}, and Tim Warburton\affilnum{3}}

\affiliation{\affilnum{1}Data Center GPU and Accelerated Processing, Advanced Micro Devices Inc., 7373 Southwest Pwky., Austin, TX, 78735.\\
\affilnum{2}Institute for Computational and Data Sciences, University at Buffalo, Buffalo, NY, 14260.\\
\affilnum{3}Department of Mathematics, Virginia Tech, McBryde Hall, 225 Stanger Street, 24061 Blacksburg, VA,  USA.}

\corrauth{Noel Chalmers\affilnum{1}.}

\email{noel.chalmers@amd.com}

\begin{abstract}
We present hipBone, an open source performance-portable proxy application for the Nek5000 (and NekRS) CFD applications.  HipBone is a fully GPU-accelerated C++ implementation of the original NekBone CPU proxy application with several novel algorithmic and implementation improvements which optimize its performance on modern fine-grain parallel GPU accelerators. Our optimizations include a conversion to store the degrees of freedom of the problem in assembled form in order to reduce the amount of data moved during the main iteration and a portable implementation of the main Poisson operator kernel. We demonstrate near-roofline performance of the operator kernel on three different modern GPU accelerators from two different vendors. We present a novel algorithm for splitting the application of the Poisson operator on GPUs which aggressively hides MPI communication required for both halo exchange and assembly. Our implementation of nearest-neighbor MPI communication then leverages several different routing algorithms and GPU-Direct RDMA capabilities, when available, which improves scalability of the benchmark.  We demonstrate the performance of hipBone on three different clusters housed at Oak Ridge National Laboratory, namely the Summit supercomputer and the Frontier early-access clusters, Spock and Crusher. Our tests demonstrate both portability across different clusters and very good scaling efficiency, especially on large problems. 
\end{abstract}


\maketitle

\section{Introduction}

\label{intro.sec}
Nek5000 \citep{nek5000} is a high-order incompressible Navier-Stokes solver used in large-scale computational fluid dynamics applications.  It is widely used in the computational fluid dynamics (CFD) research community and is commonly used on the world's largest supercomputers.  Understanding how large complex applications will perform on new and evolving architectures is made easier by distilling key computational components into proxy applications.  NekBone\footnote{\url{https://github.com/Nek5000/NekBone}} is a proxy-application for Nek5000 and only solves for the pressure term in the Navier-Stokes equations since the pressure solve is one of the most computationally intensive portions of solving the incompressible Navier-Stokes equations.  NekBone is a Tier-1 CORAL-2 benchmark\footnote{\url{https://asc.llnl.gov/coral-2-benchmarks}} and will be used to demonstrate performance on the all of the initial exascale supercomputer systems funded by the US Department of Energy CORAL-2 program.

With the growing popularity of highly heterogeneous system architectures, in particular systems with nodes containing one or more accelerators, there is a increasing need for high-performance computing (HPC) applications to leverage the often significantly higher throughput accelerators offer over traditional CPUs. Early efforts to adapt the finite element methods (FEM) to GPU accelerators were presented by \cite{goddeke2007exploring}, and \cite{goddeke2007performance}. Shortly after, both  \cite{klockner2009nodal} and \cite{komatitsch2009porting} demonstrated accelerated high-order FEM implementations using CUDA. Since these seminal works, GPUs have gradually grown in popularity for FEM applications. To name only a few, \cite{dong2014step} detail the porting of a large FEM hydrodynamics application to leverage GPU acceleration, \cite{kronbichler2017performance} and \cite{kronbichler2018architecture} and show results of using GPUs in the \verb#deal.II# FEM library, and \cite{beams2020high} detail applying GPU-accelerated batch linear algebra routines to FEM applications. 

As accelerated parallel processing becomes more ubiquitous, portability between different vendors becomes of concern as each platform may offer several different (and often competing) programming models which enable developers to extract high throughput from accelerators. Some of the most popular programming models are OpenMP, OpenACC, CUDA, HIP, OpenCL\TM, and SYCL\TM, but few of these are available/consistent across all vendors, and those that are often rely on advanced and slow-moving compiler technologies. This is especially true for Fortran applications such as Nek5000, where native options are limited to OpenMP, OpenACC, or CUDA Fortran for GPU-acceleration. 

There have been several efforts to enable GPU-acceleration in the Nek5000 application. In a series of papers, \cite{gong2014nek5000}, \cite{markidis2015openacc}, \cite{cebamanos2014auto}, \cite{otero2019openacc}, and \cite{vincent2021strong} have studied an OpenACC version of Nek5000. As is natural for a proxy application, these efforts often begin by studying the porting and performance of NekBone. \cite{gong2016nekbone} demonstrated a GPU-accelerated version of NekBone using OpenACC and CUDA Fortran. This version was later improved by \cite{karp2020optimization} using native CUDA C kernels with implementations based on the algorithms from \cite{swirydowicz2019acceleration}. Porting NekBone to FPGAs was also studied by \cite{brown2020exploring}. 

While the studies of utilizing OpenACC in NekBone and Nek5000 have shown some success, the increasing landscape of GPU-accelerated systems as part of the US Exascale initiative has made extracting even more performance from GPU accelerators, across several GPU vendors, imperative. It is therefore desirable to explore implementing Nek5000 on other programming models using some portability framework. To this end, a new GPU-accelerated version of Nek5000, namely NekRS, was presented by \cite{nekrs}. NekRS is an open-source C++ version of Nek5000, providing access to the standard Nek5000 interface and features allowing users to leverage existing application-specific source code and data files on GPU-based platforms. NekRS is built off an early fork of the libParanumal project by \cite{libparanumal2020}, a self-contained high-order finite element library that uses highly optimized kernels and leverages the Open Concurrent Compute Abstraction (OCCA) (see \cite{occa}) for performance and portability across several parallel programming models. 

In this paper we present hipBone, a C++ implementation of the main components of NekBone, which serves as a proxy for NekRS in much the same way as NekBone does for Nek5000. Like NekRS, hipBone is written using several simplified libParanumal libraries and leverages performance portability via OCCA. From the outset, we designed hipBone to be portable, to extract high performance from modern GPU accelerators, and scale to many GPUs in a compute cluster. With this design in mind, we describe several implementation and algorithmic differences between hipBone and NekBone. Notably, we choose to store the degrees of freedom of the global problem in assembled form, which significantly reduces the amount of data motion in each iteration of the benchmark at the cost of needing two distinct nearest neighbor communication phases. To counter this, we also present a novel splitting algorithm for computing the action of the main operator in hipBone which aggressively hides MPI communication overheads by local computation. We also implement several routing algorithms to perform the nearest neighbor MPI communications, and leverage GPU-Direct RDMA technologies in MPI when available. We demonstrate high performance of the main operator kernel, and then present scaling studies on some current-generation GPU accelerators to demonstrate that hipBone obtains high performance and scalability on many GPUs.  

The remainder of this paper is organized as follows. We begin by giving an overview of the algorithm used in NekBone and the computations performed. We then give an overview of the hipBone benchmark, describing the major implementation differences with NekBone, including the Poisson operator kernel, data layout, MPI nearest-neighborhood collectives, and communication hiding. We present near optimal performance of the Poisson operator kernel in hipBone on NVIDIA Tesla V100, AMD Instinct MI100, and AMD Instinct MI250X GPU accelerators. Finally, we demonstrate efficient scaling of the hipBone benchmark using the Oak Ridge National Laboratory (ORNL) Summit supercomputer, as well as the Frontier early-access Spock and Crusher clusters at ORNL, and give some concluding remarks.

\section{The NekBone Benchmark}
The \cite{nekbone} benchmark solves a screened Poisson equation discretized via the spectral element method (SEM) using a conjugate gradient iterative method. It is formulated to serve as a proxy-application for the  principal computational phase of the Nek5000 application (see: \cite{nek5000}).

The full Nek5000 application is an open-source Fortran code for simulating incompressible ﬂows. Nek5000 uses MPI for parallel communication, but otherwise has no additional threading or GPU acceleration. The discretization scheme used in Nek5000 is based on the spectral-element method (see: \cite{patera1984spectral} and \cite{tufo1999terascale}). In particular, the domain is discretized into hexahedral elements and the  high-order polynomial basis functions are chosen as the tensor-product of Lagrange polynomials, interpolating the Gauss-Legendre-Lobatto (GLL) quadrature points.

The NekBone benchmark follows a similar design to the Nek5000 application; NekBone is an open-source Fortran code, using MPI for parallelization. NekBone implements a Conjugate Gradient (CG) iterative method (listed in detail in Algorithm \ref{alg:cg}) to solve a screened Poisson equation. In particular, NekBone runs a fixed 100 iterations of the CG method, recording the total time taken, and outputting a computed floating-point operations per second (FLOPS) metric as the figure-of-merit (FOM). 

The problem setup in NekBone consists of regular mesh of $E$ hexahedral elements, together with a degree $N$ polynomial discretization amounting to $(N+1)^3$ interpolation points on each element. Collected across the entire mesh, this totals to $N_L = E(N+1)^3$ element-local interpolation points. However, as many points are shared between one or more elements along element boundaries, the actual number of degrees-of-freedom (DOFs), $N_G$, is strictly smaller than $N_L$. The exact number depends on the mesh geometry and boundary conditions, but $N_G \approx EN^3$ for large meshes. 

\begin{algorithm}[t]
  \caption{Conjugate Gradient Method}
  \label{alg:cg}
\begin{boxedminipage}{0.5\textwidth}
    \begin{algorithmic}[1]
    \STATE {\bf Input:} (1) Initial guess $\mathbf{x}$, (2) Right-hand-side vector $\mathbf{b}$, (3) Linear operator $A$, (4) Tolerance $\epsilon$.
    \STATE {{\bf Output:} (1) Solution $\mathbf{x}$, (2) Iteration count $j$}
    \STATE {Set $j=0$}
    \STATE {Set $\mathbf{r}_0 = \mathbf{b}-A\mathbf{x}$}
    \STATE {Set $\mathbf{p} = \mathbf{r}$}
    \WHILE {($\mathbf{r}_j\cdot\mathbf{r}_j > \epsilon$)}
        \STATE {$\alpha = \frac{\mathbf{r}_j\cdot\mathbf{r}_j}{\mathbf{p}\cdot A\mathbf{p}}$}
        \STATE {$\mathbf{x} = \mathbf{x} + \alpha\mathbf{p}$}
        \STATE {$\mathbf{r}_{j+1} = \mathbf{r}_j - \alpha A\mathbf{p}$}
        \STATE {$\beta = \frac{\mathbf{r}_{j+1}\cdot\mathbf{r}_{j+1}}{\mathbf{r}_j\cdot\mathbf{r}_j}$}
        \STATE {$\mathbf{p} = \mathbf{r}_{j+1} + \beta \mathbf{p}$}
        \STATE {$j=j+1$}
    \ENDWHILE
  \end{algorithmic}
\end{boxedminipage}
\end{algorithm}

The screened Poisson operator in NekBone can be written as 
\begin{equation}
A = S + \lambda I,
\label{eq:poisson}
\end{equation}
where $S$ is SEM discrete Laplacian operator, and $I$ is the identity matrix. The discrete Laplacian is derived from the typical variational form of the Laplacian. It can be written as an `assembled' action from each element, that is
\begin{equation}
S = Z^T S_L Z.
\label{eq:stiffness}
\end{equation}
While not explicitly formed, the sparse $N_L \times N_G$ matrix $Z$ in \eqref{eq:stiffness}, also called simply the `scatter' operator, is a boolean matrix containing a single non-zero per row, encoding the connectivity of the global mesh of grid-points. Its transpose, $Z^T$ is often called the `gather' operator. The reader is referred to \cite{canuto2012spectral}, \cite{deville2002high}, and \cite{henderson1995unstructured} for a more complete description of the gather and scatter operators. As these operations involves data from neighboring elements, the entirety of the MPI communication required for the SEM Poisson operator are contained in these operations.  

The $N_L \times N_L$ matrix $S_L$ in \eqref{eq:stiffness} is block-diagonal, each block $S^e_L$ corresponding to element $e$ and can be written as a product of several operators:
\[
S_L^e =  \mathbf{D}^T \mathbf{G}^e \mathbf{D}.
\]
Here, $\mathbf{D}$ is a $3(N+1)^3 \times (N+1)^3$ discrete gradient operator which can be written in the following tensor-product form
\[
\mathbf{D} = \begin{pmatrix} D \otimes I \otimes I \\ I \otimes D \otimes I \\ I \otimes I \otimes D \end{pmatrix},
\]
where $D$ is the $(N+1)\times (N+1)$ one-dimensional SEM derivative operator, and $I$ is the identity matrix. The metric tensor geometric factor, $\mathbf{G}^e$, is a $3(N+1)^3 \times 3(N+1)^3$ matrix which has the following symmetric form:
\begin{equation*}
\mathbf{G}^{e} = \begin{pmatrix}
G^e_{rr} & G^e_{rs} & G^e_{rt} \\
G^e_{rs} & G^e_{ss} & G^e_{st} \\
G^e_{rt} & G^e_{st} & G^e_{tt} 
\end{pmatrix},    
\end{equation*}
where each $G^e$ is a diagonal $(N+1)^3 \times (N+1)^3$ matrix of pointwise values of the corresponding entry of the element's metric tensor, combined with GLL quadrature weights.

The tensor product form allows the local Poisson operators, $S^e_L$, to be implemented using highly-tuned matrix-matrix multiplication routines in the NekBone benchmark. Since each multiplication by the one-dimensional derivative matrix $D$ requires $2(N+1)^2$ FLOPs, the application of the full gradient operator $\mathbf{D}$ and its transpose $\mathbf{D}^T$ require $12(N+1)^4$ FLOPs per element. On the other hand, the multiplication of the geometric factors $\mathbf{G}^e$ requires only $15(N+1)^3$ FLOPs per element.

Rather than storing the length $N_G$ DOF vector $\mathbf{x}_G$, NekBone, like Nek5000, instead stores the scattered vector $\mathbf{x}_L = Z\mathbf{x}_G$, which has length $N_L$. This requires extra memory, since $N_L>N_G$, but has the effect that computing the action of the screened Poisson operator in \eqref{eq:poisson} can be written as
\begin{align*}
 \mathbf{b}_G &= A\mathbf{x}_G, \\
\Rightarrow  Z\mathbf{b}_G &= Z(Z^TS_LZ + \lambda I)\mathbf{x}_G, \\
Z\mathbf{b}_G &= ZZ^TS_LZ\mathbf{x}_G + \lambda Z\mathbf{x}_G, \\
\mathbf{b}_L &= (ZZ^T S_L + \lambda I)\mathbf{x}_L.
\end{align*}
That is, the action of the discrete Laplacian, $S$, can be computed by applying the element-local $S_L$ to $\mathbf{x}_L$, followed by the `gather-scatter' operator, $ZZ^T$. Since all MPI communication is contained inside this operation, a combined gather-scatter routine is implemented and communication is done using custom exchange algorithms implemented inside the \verb.gslib. library, based on the work of \cite{tufo1998algorithms} and \cite{fischer2008petascale}. An in-depth study of the performance of specialized communication routines in NekBone on structured grids was present by \cite{ivanov2015evaluation}.

Storing the DOF vector in the scattered form leads to a subtlety in the CG iteration. Since the scattered vectors have repeated values whenever degrees of freedom are shared between elements, computation of global inner products as part of the CG method therefore requires a `weighted' variant, where each contribution to the global summation is weighted by the inverse counting vector (i.e. the entries of the diagonal matrix $Z^TZ$). This weight vector is pre-computed and of length $E(N+1)^3$, but the computation of inner product does require reading this vector from memory and performing $E(N+1)^3$ additional FLOPs. 

Taking the weighted inner products into account,  the total FLOP count per CG iteration is the sum of the total FLOP count for the evaluation of the screen Poisson operator, together with the assortment of vector operations as part of the CG iteration. The NekBone benchmark uses the total of 
\begin{equation}\label{eq:nekbone_flops}
12E(N+1)^4 + 34E(N+1)^3,    
\end{equation}
FLOPs per CG iteration in order to compute its FLOPS figure-of-merit.

\section{HipBone}
HipBone, our re-implementation of the NekBone benchmark, is built upon a core set of libraries from the open-source finite element library libParanumal (cf. \cite{libparanumal2020}). These core libraries provide routines for simple hexahedral mesh generation, finite element operator construction, accelerated linear algebra and linear solvers, and nearest-neighbor communication collectives. 

As part of the libParanumal core, hipBone uses the Open Concurrent Compute Abstraction (OCCA) library (cf. \cite{occa}) to abstract between different parallel programming models such as OpenMP, OpenCL\TM, CUDA, HIP, and SYCL\TM. OCCA is a required dependency of libParanumal and provides a portable abstraction through a unified API which allows users to implement parallel kernel code in a (slightly decorated) C++ language, called OKL. At runtime, the user can specify which parallel programming model to target. OCCA then translates the OKL source code into the desired target language and Just-In-Time (JIT) compiles kernels for the user's target hardware architecture. Leveraging JIT compilation also allows different vendor compilers additional optimization, e.g. loop unrolling.

The hipBone benchmark itself is fairly lightweight on top of the libParanumal libraries. In addition to simple vector operator kernels, the benchmark contains only two additional kernels. The first is used only once to populate a pseudo-random initial forcing vector, and the second is the performance-critical screened Poisson operator. 

HipBone does not replicate the full functionality of NekBone at this point in time. Most notably, hipBone does no form of preconditioning in the CG iteration, while NekBone supports simple diagonal preconditioning as well as a much more sophisticated hybrid-Schwarz multigrid preconditioner detailed in \cite{lottes2005hybrid}. We omit this functionality for now in order to first focus on optimized GPU performance, portability, and scalability of the Poisson operator and streaming operations in hipBone, leaving the topic of preconditioning on modern GPU accelerators for future study. There are, additionally, several key algorithmic/implementation differences between hipBone and NekBone. In the remainder of this section, we overview the most significant changes and their impact on performance. 

\subsection{DOF Storage} \label{dof.sec}
As described above, the NekBone benchmark stores the DOF vectors in scattered form, with $N_L$ entries instead of $N_G$. This implementation choice has the benefit of allowing the MPI communication required in the Laplacian operator to be combined into a single gather-scatter operation, reducing MPI messaging latency costs. This is especially beneficial when strong-scaling a fixed problem size to many MPI ranks. The trade-off, however, is that vectors are now (potentially significantly) longer. Indeed, $\frac{N_L}{N_G} \approx \frac{(N+1)^3}{N^3}$, which is a still a 21.3\% overhead at the relatively high order of $N=15$. Furthermore, global reduction operations such as inner products require accessing an additional weight vector, leading to even more required data movement during each CG iteration. 

Each of the vector operations in the CG iteration are of very low arithmetic intensity, i.e. they perform very little (often less than one) FLOPs per byte of data loaded from main memory. Even the element-local Laplacian operator has a fairly low arithmetic intensity, owing to the tensor-product form. Such low arithmetic intensity operations are often referred to as `streaming' operations, and their runtimes $t$ are well-modeled by an Amdhal-like model $t= \alpha + \beta B$, where $\alpha$ is some fixed latency cost, $\beta$ is an effective asymptotic streaming bandwidth, and $B$ is the amount of data moved in the operation. The reader is referred to \cite{hockney1982characterization} and \cite{hockney1985r} for a more complete discussion, and more recently \cite{chalmers2020portable} who studied streaming performance on several modern GPU-accelerators. With GPUs, the $\alpha$ term includes the latency cost of off-loading a kernel to the device and eventually synchronizing back with the host process. This is often referred to as a kind of `kernel launch latency'. However, once a significant amount of data is moved in HBM (typically $\sim$10MB), performance becomes dominated by the effective HBM bandwidth of the accelerator, $\beta$. 

When targeting modern GPU-accelerators, the implementation choice of storing vectors in scattered form in NekBone has the effect of increasing data movement (therefore reducing effective performance) for large problem sizes. On the other hand, it also reduces MPI latency costs for small problem sizes, but in this regime the GPU's performance is primarily limited by kernel launch latency. With hipBone, we are interested in the case where the problem size is at least large enough to saturate the main memory bandwidth of the GPU. We therefore have made the implementation choice to store the DOF vectors in their assembled ordering. With this choice, vectors in the CG iteration become length $N_G$, rather than $N_L$, and some data motion, such as the accessing the weight vector in the global inner products, is avoided completely. This allows for significant effective performance improvements for large problems in each of the vector operations in the CG iteration. This design choice also aligns closely with the implementation of libParanumal's elliptic solvers, and the MFEM finite element library (cf. \cite{anderson2021mfem}). 

\subsection{Poisson Operator}
The performance of the SEM Laplacian operator on hexahedral elements on modern GPU-accelerators has been studied previous by several authors.  \cite{remacle2016gpu} presented an implementation of a high-order spectral element elliptic solver on GPU accelerators. In this work, the authors detail a Laplacian kernel that exploits fine-grain parallelism of the accelerator by mapping each of the elements to distinct threadblocks, and assigning each DOF in an element to an individual GPU thread. This approach creates a 3D thread structure of $(N+1)^3$ in each threadblock. This approach was limited to $N=9$ due to the 1024 thread-per-block limit in CUDA, and $N=5$ due to OpenCL's limit of 256 work-items per work-group on AMD GPUs. 

The approach of \cite{remacle2016gpu} was later extended by \cite{swirydowicz2019acceleration} as part of the Center for Efficient Exascale Discretizations (CEED) Exascale Computing Project co-design center\footnote{\url{http://ceed.exascaleproject.org}}. In this work, the authors studied the GPU-implementation and optimization of several high-order finite element operators, including mass matrix multiplication, and diffusion matrix multiplication with/without higher-order quadrature rules. The case of the diffusion operator without higher-order quadrature is particularly useful for our hipBone implementation, as this aligns very closely with the screened Poisson operator in NekBone. The authors detail two optimized kernel implementations. The first follows the approach of \cite{remacle2016gpu} by using a 3D threadblock structure, but has additional optimizations such as processing several elements in a single threadblock for small polynomial orders, $N$. The second implementation uses a layer-by-layer processing of the hexahedral element, mapping individual GPU threads to $(N+1)^2$ nodes of a single face of the hexahedral (see for instance \cite{stilwell2013gnek}). This implementation makes heavy use of register space, but allows for higher polynomial order $N$ to be used. Both implementations heavily rely on shared memory as a scratchpad to store temporaries, and many of the FLOPs performed in the kernels have one or more of the inputs/outputs residing in shared memory. This can become a performance limiter as arithmetic intensity increases, as loading data to/from shared memory can be significantly slower than performing the FLOP. 

In hipBone, we implement a single kernel to compute an intermediate vector $\mathbf{y}_L = (S_L+\lambda W)Z\mathbf{x}_G$, where $W$ is a diagonal matrix of the inverse degree weights which can be defined as the diagonal matrix which satisfies $Z^TW = I$. We compute this intermediate vector $\mathbf{y}_L$ so that the final action of the Poisson operator $A$ can be computed by gathering $\mathbf{y}_L$, i.e. $A\mathbf{x}_G = Z^T\mathbf{y}_L$, which requires MPI communication. To compute $\mathbf{y}_L$ we use a combined approach of the 3D and 2D threadblock structure kernels from \cite{swirydowicz2019acceleration}, with some minor changes. First, for $N<9$ we use the 3D threadblock kernel, while for larger $N$ we use the 2D threadblock structure kernel. The decision to use the 2D threadblock kernel for $N=9$ is made based on empirical performance measurements and current generation GPU accelerators, but is configurable. We also implement a blocking strategy in both kernel implementations, which allows for multiple elements to be processes by a single threadblock. This helps to minimize idle threads in warps/wavefronts. Finally, we slightly alter the storage of the geometric factors, $G^e$, so that all geometric factors at a given degree of freedom in an element are packed together. When SIMD lanes of a processor subsequently load the first geometric factor, the load will be strided since each lane loads factors different degrees of freedom. However, this load populates the lowest-level cache, and the packed storage format makes the loads of the remaining geometric factors more efficient.

The fused action of $S_L+\lambda W$ and $Z$ into a single kernel is a notable difference in our implementation compared to that presented by \cite{swirydowicz2019acceleration}. This fusing causes the reading of the element-local DOFs into the local memory on the processor to be an indirect read of $\mathbf{x}_G$. This requires additional indexing data to be stored in memory and read-in by the kernel before accessing entries of $\mathbf{x}_G$, but the ordering of elements in the mesh and repetition of DOFs shared between elements can allow for some entries of $\mathbf{x}_G$ to be found in cache during this indirect load. With MPI parallelism, this indirect read also relies on a halo region to be populated with DOFs belonging to other processes. We detail this procedure, and the MPI parallelization of the gather operator $Z^T$, below.  

The total amount of data moved through main memory and the total FLOPs done by the Poisson operator can be estimated as follows. First, the load of $\mathbf{x}_G$ into element-local storage requires a 4 byte index, encoding a row of the scatter operator $Z$, to load each 8 byte entry of $\mathbf{x}_L$, totaling to $12N_L$ bytes loaded. With perfect caching of repeatedly used entries of $\mathbf{x}_G$, the amount of data moved from main memory can actually be as low as $8N_G + 4N_L$. The load of the six geometric factors in $G^e$ and inverse degree count $W$ for each of the $N_L$ entries total to an additional $56 N_L$ bytes, and the write of the final $8N_L$ bytes for the output of $\mathbf{y}_L$ brings the total data motion to approximately $8N_G + 68 N_L \approx 8EN^3 + 68 E(N+1)^3$ bytes, assuming perfect caching. The FLOP count, on the other hand, consists of $12E(N+1)^4 + 15E(N+1)^3$ FLOPS to apply $S_L$ and $3E(N+1)^3$ additional FLOPS to add the $\lambda W$ contribution, for a total of $12E(N+1)^4 + 18E(N+1)^3$ FLOPS.

With these estimates for data motion and FLOP count, we model the compute rate, $R$, of the screened Poisson operator kernel, in FLOPS, using a typical roofline model:
\begin{equation}\label{eq:op_perf}
    R = \min\left(C, \frac{12(N+1)^4 + 18(N+1)^3}{8N^3 + 68 (N+1)^3}B\right),
\end{equation}
where $C$ is the peak compute rate of the processor in FLOPS, and $B$ is the peak (bidirectional) bandwidth of the processor's main memory in bytes/s. As most current-generation GPU accelerators have peak FP64 compute rates typically measured in TFLOPS, and peak HBM memory bandwidths near 1 TB/s, we see that the low arithmetic intensity of the Poisson operator implies performance will be primarily limited by memory bandwidth, even for large degrees $N$. We demonstrate this on several accelerators in the computational tests below. 

\subsection{MPI Communication} \label{mpi.sec}
The global mesh of hexahedral elements in hipBone is parallelized over multiple processes by partitioning the mesh evenly among the $P$ processes. Message passing between process is then done via MPI, with a given processes requiring communication from processes that border it along a face, edge, or corner of an element. While the mesh used in NekBone and hipBone is a structured cube of uniform hexahedrals, the message passing algorithms used assume no underlying mesh structure. This more accurately represents the larger Nek5000 and NekRS codes, which are generally unstructured. 

In hipBone, we have re-written and extended the functionality of the \verb.gslib. library, used in NekBone and Nek5000 for MPI communication, to a `device-aware' gather-scatter library. This new library replicates the combined gather-scatter operations provided by \verb.gslib., and as well as its various communication algorithms, on both host memory or OCCA device buffers. The library also provides new gather, scatter, and halo exchange operations, which are needed for the screened Poisson operator in the assembled DOF ordering. Since each of these MPI communication operations in the new library can cast as a sparse nearest-neighbor collective operation, and the underlying communication algorithms from the gather-scatter operation can be re-used for other operations. The library also supports leveraging GPU-Direct RDMA capabilities by directly passing pointers to device memory to MPI calls when the MPI library being used is ``GPU-aware". This allows the MPI implementation to leverage high bandwidth GPU-to-GPU or GPU-network links in a system when they are present.

Our new device-aware gather-scatter library implements the following exchange routines for nearest neighbor collective communication:

\subsubsection*{All-to-all}
The simplest of the exchange routines, the All-to-all exchange performs all required data movement between via a single \verb+MPI_Alltoallv+ function. This exchange heavily relies on existing optimized sparse all-to-all communication algorithms in the MPI implementation. Since this MPI function is one of most general MPI collective patterns, the performance of this algorithm is typically not optimal.

\subsubsection*{Pairwise}
Another relatively simple exchange algorithm, the pairwise exchange implements the nearest neighbor collective operation by simply having every process send/receive all needed data to/from its neighbors, using several calls to \verb+MPI_Isend+ and \verb+MPI_Irecv+. This exchange algorithm communicates using the maximum number of MPI messages, thereby incurring the maximum amount of messaging latency, but with direct routing moves the least amount of data possible. For very large problems where the inter-process exchanges are sensitive to the bandwidth of the network and not messaging latency, this exchange likely performs well. 

\subsubsection*{Crystal Router}
The crystal router algorithm performs the nearest neighbor collective via recursive hypercube folding. The algorithm is described in full by \cite{lamb1988solving}, and some performance results of this algorithm's use in Nek5000 were presented by \cite{schliephake2014performance}. For a $P$ process grid, the algorithm can be summarized as follows: divide the grid in half and pair each process $p_l$ in the lower half with a distinct process $p_h$ in the upper half. Then, if process $p_l$ has data needed by \textit{any} process in the upper half, it sends that data to $p_h$, and vice versa for its partner $p_h$. The algorithm then proceeds recursively on the two halves of the grid. 

The crystal router performs the nearest neighbor collective in $\lceil \log_2(P) \rceil$ bidirectional messages, thereby minimizing the total number of messages sent/received. With a suitable distribution of the global mesh, such as that obtained by recursive mesh bisection, some messages may be avoided altogether. The amount of data sent and received in total, however, is larger than the pairwise exchange. This exchange routine, therefore, is primarily useful for smaller problems that are sensitive to network latency. 

\subsubsection*{}
During the initial setup of the gather-scatter library, each of the exchange routines is timed, and the fastest exchange is selected for use in subsequent communication operations. When exchanging device-resident data, the gather-scatter library must either move the needed data to host memory, after which the host calls the MPI routines, or use GPU-aware MPI routines where available. In either case, a buffer of the data needed for communication is first extracted from the input vector using a device kernel. After this kernel completes, the user is free to queue additional work to the device to hide some or all of the time spent in MPI routines. Once the communication is complete on the host or device, the communication buffer is scattered back to its original order in the input vector. The exchanges, in addition to the MPI messaging latency costs, also incur device synchronization and kernel launch latency costs during the exchange as the host process must synchronize with the device to ensure the communication buffer is ready to be sent with MPI. When performing a crystal router exchange in device memory with GPU-aware MPI, we incur $\lceil \log_2(P) \rceil$ total device synchronizations and kernel launch latencies, as intermediate kernels are required to prepare the data to be sent in the next hypercube fold. As this exchange is primarily useful for small, latency-sensitive problems, it is unlikely for the device-resident crystal router to be the fastest exchange. 


\subsection{Overlapping halo and gather communication}

\begin{figure}
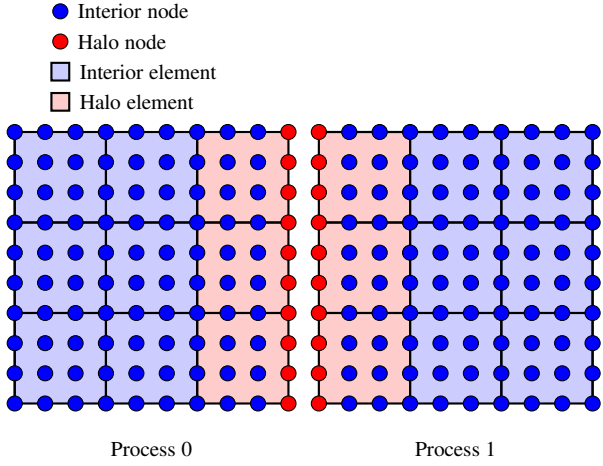

    \centering
    \ctikzfig{element}
    \caption{A two MPI process mesh arrangement of third-order 2D spectral elements.}
    \label{fig:element}
\end{figure}

At a high level, the Poisson operator application in hipBone is broken down into three stages: halo data communication, element-local Poisson operator application, and finally a gather operation. The first and third stages both require MPI communication, thus any efforts to hide this communication by local computation will greatly improve scaling efficiency.  To describe how such communications are hidden, it is useful to first define a few terms relating to the degrees of freedom in elements owned by an arbitrary MPI process (Figure \ref{fig:element}).

An arbitrary MPI process has `ownership' of a number of spectral elements, meaning that in the scattered DOF storage each element, i.e. its constituent nodes, is stored entirely on a single distinct MPI rank. At the inter-process boundary between elements in the global mesh are nodes which are contained in two or more elements owned by different processes. We call these nodes, `halo nodes', and all other nodes we call `interior nodes'. `Halo elements' are subsequently defined as elements which contain at least one halo node and `interior elements' are all elements that are not halo elements.  In Figure \ref{fig:element}, we depict this for the case of two processes in two dimensions for simplicity. In the figure, each process owns nine quadrilateral elements with a degree $N=3$ grid of interpolation nodes, and the inter-process boundary between them leads to several halo nodes in six halo elements. On each process, each of red-shaded elements is a halo element and the six remaining blue-shaded elements are interior elements.  Within each halo element is a mixture of interior and halo nodes. In Figure \ref{fig:element}, the halo nodes are colored red and the interior nodes are colored blue.

While there is a well-defined notion of `ownership' of an element by a process, in the assembled DOF ordering it is less clear what process should `own' the degree of freedom associated to a halo node. In hipBone, when defining the assembled DOF order, the owner of each halo node is chosen randomly, but fairly, from among the owners of the halo elements of which this node is a part. Consequently, when collecting the assembled DOFs into element-local form via the scatter operator $Z$, some DOFs must first be communicated via a halo exchange. The first step in the application of the screened Poisson operator in hipBone is to launch a kernel that extracts and packs buffers of halo node data to be communicated over MPI. Once complete, a kernel which computes the action of $(S_L+\lambda W)Z$ on \textit{half of the interior elements} is launched. While this kernel is executing, halo data is communicated over MPI and packed into a halo region on the destination processes. Once complete, a kernel is launched to apply the action of $(S_L+\lambda W)Z$ to all the halo elements.

\begin{figure}
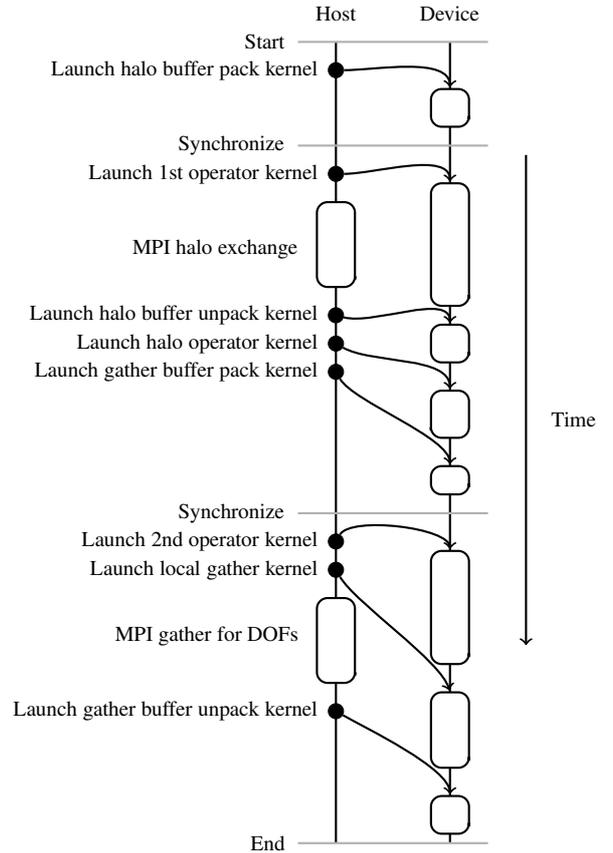

    \centering
    \ctikzfig{operator_timeline}
    \caption{A timeline visualization of the operator application logic in hipBone.  The illustration shows the splitting of the operator application into three kernels. Two kernels apply the operator to distinct halves of the interior elements, and the third kernel applies the operator to the halo element.  This strategy allows hipBone to maximize the potential for hiding MPI communications for both the halo exchange and subsequent gather operation.}
    \label{fig:operator_timeline}
\end{figure}

At this point, all that is left to do is to apply the operator to the other half of the interior elements, and then gather the intermediate vector $\mathbf{y}_L$.  The interior nodes do not need to send any data to other process for the gather operation, and so the strategy for hiding MPI communication in the gather is similar to the strategy used for hiding the halo data communication. Following the kernel which computes the local operator application on the halo elements, we queue a kernel to gather values at the halo nodes into MPI buffers to be communicated as part of the global action of $Z^T$. Once the buffers are assembled, a kernel is launched to compute the action $(S_L+\lambda W)Z$ on the remaining half of the interior elements. Another kernel, which computes the entirely process-local action of the gather operator $Z^T$ is also queued at this point. While both of these kernels are executing, the MPI communication required for the gather operator is performed, and the resulting gathered values at the halo nodes is written to the output vector. We show a graphical representation of this procedure in Figure \ref{fig:operator_timeline}. The figure illustrates a possible timeline of host and device activity during the application of the Poisson operator.

\subsection{Conjugate gradient Iteration}

The remaining operations in the Conjugate Gradient method are relatively simple. After computing $A \mathbf{p}$ for the current $\mathbf{p}$ vector in line 7 of Algorithm \ref{alg:cg}, the operations for the remainder of the iteration are each simple vector operations. The only remaining MPI communication required comes with the computation of global inner products $\mathbf{r}\cdot\mathbf{r}$, and $\mathbf{p}\cdot A\mathbf{p}$. Implementing these global reductions along with the rest of the vector operations in the CG iteration offered some additional opportunities for optimization.

We assume that at the beginning of the $j$-th iteration we have already computed the current value of $\mathbf{r}_j\cdot\mathbf{r}_j$. The computation of $\alpha$ in line 7 of Algorithm \ref{alg:cg} only requires the inner product $\mathbf{p}\cdot A\mathbf{p}$, which we locally compute with a dedicated inner product kernel on each rank, and complete with a \verb+MPI_Allreduce+. Using $\alpha$, we compute $\mathbf{r}_{j+1}$ and simultaneously accumulate the local inner product $\mathbf{r}_{j+1}\cdot\mathbf{r}_{j+1}$. Fusing this reduction with the update of $\mathbf{r}$ avoids the need for a separate kernel to read the vector $\mathbf{r}$ again. Moreover, the subsequent \verb+MPI_Allreduce+ call which completes the global reduction can be well-hidden by first queuing a kernel which performs the AXPY update of $\mathbf{x}$ in line 8 of Algorithm \ref{alg:cg} beforehand. This optimization slightly improves scaling efficiency by hiding the latency-sensitive MPI collective communication time. 

For the entire CG iteration, recall from the discussion above that the screened Poisson operator kernel moves a total of $8N_G+ 68N_L$ bytes through main memory. The gather operator $Z^T$ reads a vector of length $N_L$, along with a compressed sparse-row (CSR) encoding of the non-zero structure of $Z^T$, and outputs a vector of length $N_G$ totaling to $12N_G + 12N_L$ bytes moved through main memory. The remaining vector operations; the inner product, fused AXPY and inner product, and two AXPYs, total to 11 more reads and writes of vectors from main memory, totaling $88 N_G$ bytes. This brings the total data motion to 
\begin{align*}
& 108 N_G + 80 N_L, \\
\approx & 108 EN^3 + 80 E(N+1)^3,    
\end{align*}
bytes moved per CG iteration. Similarly, the number of FLOPs done by the screened Poisson kernel is $12E(N+1)^4 + 18E(N+1)^3$, the subsequent gather performs $E(N+1)^3$ FLOPs, and the remaining vector operations require only $10 N_G$ FLOPs, bringing the total FLOPs to approximately
\begin{equation}\label{eq:hipbone_flops}
12E(N+1)^4 + 19E(N+1)^3 + 10 EN^3,    
\end{equation}
FLOPs per CG iteration. Comparing this to the FLOP count \eqref{eq:nekbone_flops} used in the original NekBone benchmark, hipBone's FLOP count has the same leading order term, $12E(N+1)^4$, from the screened Poisson operator application, but performs slightly fewer FLOPs overall due to the unweighted inner products and smaller vector sizes. This difference has only a minor impact on the reported FOM in FLOPS. For consistency with other NekBone studies, we use the original FLOP count from NekBone when reporting the FOM in our numerical experiments below. 

\section{Computational Tests}
\label{tests.sec}

This section describes the computational studies we performed on three different machines at the Oak Ridge Leadership Compute Facility (OLCF): Summit, Spock, and Crusher.  Summit\footnote{\url{https://docs.olcf.ornl.gov/systems/summit_user_guide.html}} is an IBM system,  each node consisting of a dual socket configuration of two IBM POWER9 CPUs, six NVIDIA Tesla V100 GPUs, and one dual-rail 100Gbps Mellanox Connect X-5 EDR InifniBand network card. Spock\footnote{\url{https://docs.olcf.ornl.gov/systems/spock_quick_start_guide.html}} is an HPE system, each node consisting of a single socket AMD EPYC 7662 CPU, four AMD Instinct MI100 GPUs, and one HPE Slingshot 100Gbps network card. Finally, Crusher\footnote{\url{https://docs.olcf.ornl.gov/systems/crusher_quick_start_guide.html}} is an HPE Cray EX supercomputer system serving as a Frontier early-access system,  each node consisting of a single socket optimized 3rd Gen EPYC 64 core processor, four AMD Instinct MI250X accelerators, and four HPE Slingshot 200Gbps network interfaces. 

On Summit, we use GCC v9.1.0 as our C++ compiler, CUDA v11.0.3 to JIT compile device kernels, and Spectrum-MPI v10.4.0.3 as the MPI implementation. On Spock, we use GCC v11.2.0 as our C++ compiler, ROCm v4.5.0 to JIT compile HIP kernels, and Cray-MPICH v8.1.12 as the MPI implementation. Finally, on Crusher we GCC v11.2.0 as our C++ compiler, ROCm v4.5.2 to JIT compile HIP kernels, and Cray-MPICH v8.1.12 as the MPI implementation. For each of these systems, we perform several performance and scaling studies and present their results below.

\subsection{Poisson Operator Performance}
Our first test measures the performance of the screened Poisson operator kernel on each of the three different GPU accelerators considered, as its performance is critical to the overall performance of the benchmark. To isolate the performance on each accelerator, we use a single NVIDIA Tesla V100, a single AMD MI100, and one of the two Graphics Compute Die (GCD) on the AMD Instinct MI250X Multi-Chiplet Module (MCM). It is important to note that while each GCD in the MI250X presents as a distinct GPU to the operating system, the two GCDs on the MI250X share the total power budget on the full MI250X MCM. Testing on a single GCD, therefore, allows for the GCD to potentially use more than half of the overall power budget of the MI250X, and operate at high clock frequencies than one would observe when running the same workload on both GCDs simultaneously. For the operator tests in this section, however, in which performance is predominately bound by HBM streaming rates, we have not observed any significant performance difference between kernels executed on a single GCD of the MI250X, or both GCDs in the MCM simultaneously. 

\begin{figure}
\centering
    \begin{subfigure}[t]{0.45\textwidth}
        \centering
        \begin{tikzpicture}[scale=0.7]
\begin{axis}[
  grid=both, 
  major grid style={line width=.1pt,draw=gray!50}, 
  minor grid style={line width=.1pt,draw=gray!50}, 
  domain=1:8, 
  width=4in, 
  ymin=1e-16, 
  xlabel={Polynomial Degree, $N$}, 
  ylabel={Poisson Operator GFLOPS}, 
  cycle list name=will, 
  legend cell align=left, 
  legend pos=north west, 
  mark size=1.5pt, 
  line width=1.4pt, 
 legend entries={Empirical Roofline, Measured Performance},
  title={NVIDIA Tesla V100},
  ymax=3500,
  ymin=0.0,
  xmin=1,
  xmax=15,
]

\addplot [domain=1:16, mark=none, thick ]
            { 820*((12*(x+1)+18)*(x+1)^3)/((68*(x+1)^3 +8*(x)^3)};
            
\addplot   table {
1 486.2
2  597.3
3  706.4
4  816.1
5  955.0
6  1082.7
7  1128.6
8  1204.0
9  1453.6
10  1435.9
11  1679.8
12  1658.1
13  1616.8
14  1808.5
15  2101.4
};

\end{axis}
\end{tikzpicture}
        \caption{Performance of Poisson operator in hipBone on a single NVIDIA Tesla V100 shown with empirical roofline model. Roofline model uses an empirical peak streaming rate of $B=820$ GB/s.}
        \label{V100Operator.fig}
    \end{subfigure}
    \hfill
    \begin{subfigure}[t]{0.45\textwidth}
        \centering
        \begin{tikzpicture}[scale=0.7]
\begin{axis}[
  grid=both, 
  major grid style={line width=.1pt,draw=gray!50}, 
  minor grid style={line width=.1pt,draw=gray!50}, 
  domain=1:8, 
  width=4in, 
  ymin=1e-16, 
  xlabel={Polynomial Degree, $N$}, 
  ylabel={Poisson Operator GFLOPS}, 
  cycle list name=will, 
  legend cell align=left, 
  legend pos=north west, 
  mark size=1.5pt, 
  line width=1.4pt, 
  legend entries={Empirical Roofline, Measured Performance},
  title={AMD Instinct MI100},
  ymax=3500,
  ymin=0.0,
  xmin=1,
  xmax=15,
]

\addplot [domain=1:16, mark=none, thick ]
            { 982*((12*(x+1)+18)*(x+1)^3)/((68*(x+1)^3 +8*(x)^3)};
            
\addplot   table {
1  596.7
2  734.7
3  882.3
4  1067.4
5  1210.4
6  1296.3
7  1469.7
8  1535.1
9  1597.2
10 1792.3
11 1573.9
12 1756.4
13 1728.4
14 1868.1
15 2135.2
};

\end{axis}
\end{tikzpicture}
        \caption{Performance of Poisson operator in hipBone on a single AMD Instinct MI100 shown with empirical roofline model. Roofline model uses an empirical peak streaming rate of $B=952$ GB/s.}
        \label{MI100Operator.fig}
    \end{subfigure}
    \hfill
    \begin{subfigure}[t]{0.45\textwidth}
        \centering
        \begin{tikzpicture}[scale=0.7]
\begin{axis}[
  grid=both, 
  major grid style={line width=.1pt,draw=gray!50}, 
  minor grid style={line width=.1pt,draw=gray!50}, 
  domain=1:8, 
  width=4in, 
  ymin=1e-16, 
  xlabel={Polynomial Degree, $N$}, 
  ylabel={Poisson Operator GFLOPS}, 
  cycle list name=will, 
  legend cell align=left, 
  legend pos=north west, 
  mark size=1.5pt, 
  line width=1.4pt, 
  legend entries={Empirical Roofline, Measured Performance},
  title={AMD Instinct MI250X, Single Die},
  ymax=3500,
  ymin=0.0,
  xmin=1,
  xmax=15,
]
            
\addplot [domain=1:16, mark=none, thick ]
            { 1117*((12*(x+1)+18)*(x+1)^3)/((68*(x+1)^3 +8*(x)^3)};
            
\addplot   table {
1 653.1
2 760.6
3 985.2
4 1101.8
5 1290.1
6 1440.5
7 1628.3
8 1727.7
9 1909.9
10 2083.8
11 2075.2
12 2280.5
13 2275.4
14 2559.1
15 2774.9
};

\end{axis}
\end{tikzpicture}
        \caption{Performance of Poisson operator in hipBone on a single GCD of AMD Instinct MI250X shown with empirical roofline model. Roofline model uses an empirical peak streaming rate of $B=1117$ GB/s.}
    \label{MI250XOperator.fig}
    \end{subfigure}
    \hfill
    \caption{Performance test of Poisson operator in hipBone for polynomial degrees $N=1,\ldots,15$. Results for each GPU accelerator are shown with empirically calibrated roofline model given in \eqref{eq:op_perf}.} 
    \label{fig:AxOperator}
\end{figure}
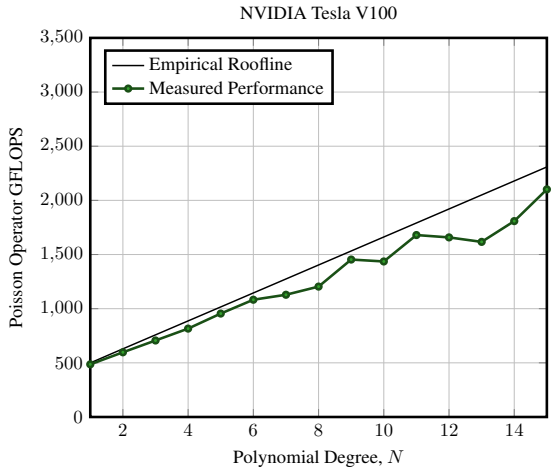
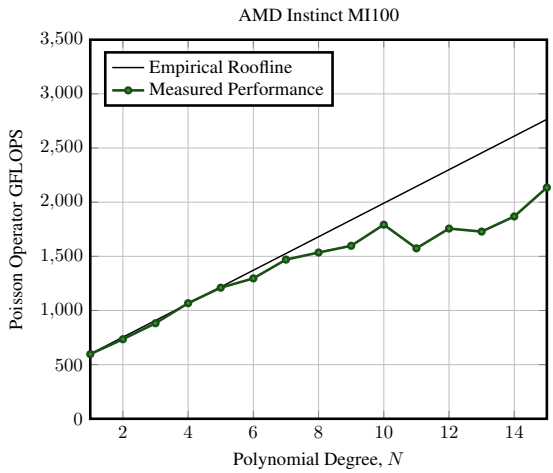
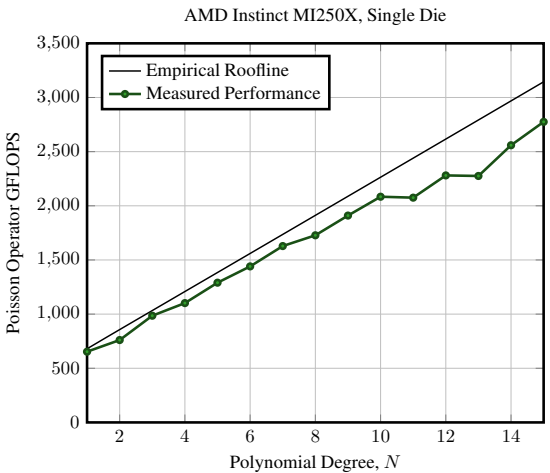

We measure the performance of the Poisson operator kernel by time-averaging 50 back-to-back invocations of the kernel for each polynomial degree $N=1,\ldots,15$. For each degree, we use a fixed mesh of elements, chosen large enough so that the number of DOFs in the mesh is approximately $N_G\approx 40$ million. This helps to minimize effects such as kernel launch latency, threadblock scheduling latency and tail effects, and core clock variations due to temporary boosting. 

We also compute, for each device, an empirically calibrated roofline using the model in equation \eqref{eq:op_perf}. We use as the peak bandwidth, $B$ in the roofline model, not the quoted theoretical peak of each device's HBM memory, but rather an empirically measured peak streaming rate. To measure this rate, we use a simple streaming kernel in which each thread in the threadblocks only reads 8 FP64 values from main memory, and writes back one FP64 value. This 8:1 read-write ratio closely matches the data movement of the Poisson operator kernel, but is also idealized in that there is no unstructured load, no data re-use in cache, and the kernel uses a uniform threadblock size. We compute an asymptotic streaming rate for this kernel using the methodology described in \cite{chalmers2020portable}, finding streaming rates of $B=810$ GB/s for the NVIDIA Tesla V100, $B=982$ GB/s for the AMD Instinct MI100, and $B=1117$ GB/s for a single GCD of the AMD Instinct MI250X. We do not treat these streaming rates, nor the roofline model itself, as absolute bounds on the performance of the Poisson operator kernel. Rather, we use this roofline model as guidance for what one would expect the performance of the operator kernel to be, if it were to read and write data to main memory as efficiently as a pure streaming kernel.

We show the results of the Poisson operator performance test in Figure \ref{fig:AxOperator}. We show the performance in GFLOPS of the operator kernel for each $N=1,\ldots,15$ on the NVIDIA Tesla V100, AMD Instinct MI100, and a single GCD of the AMD Instinct MI250X in Figures \ref{V100Operator.fig}, \ref{MI100Operator.fig}, and \ref{MI250XOperator.fig}, respectively. For each device, we see that the measured operator performance closely matches the empirical roofline model for $N<9$, indicating near-optimal performance of the 3D threadblock kernel described above. At $N=9$ and beyond, wherein the kernel implementation switches to the 2D threadblock algorithm, performance becomes more variable, with some degrees falling significantly below the roofline. This is especially noticeable in the performance on the AMD Instinct MI100 which sees a performance dip at $N=11$. While performance continues to climb after $N=11$, the gap between the measured performance and the roofline is never fully recovered. Despite the departure from the roofline model on each GPU, all the devices achieve the highest GFLOPS at the highest polynomial degree $N=15$, with the NVIDIA Tesla V100 achieving 2101.4 GFLOPS, the AMD Insitinct MI100 achieving 2135.2 GFLOPS, and a single GCD of the AMD Instinct MI250X achieving 2774.9 GFLOPS. 

The inefficiencies that appear in the Poisson operator kernel for large $N$ may be due to the combination of high register pressure, masked threads, and large threadblocks. To gain some insight into these effects, we can consider some important metrics regarding the occupancy of the Poisson operator kernel. For each polynomial degree, $N$, the threadblock size of the 2D threadblock operator kernel is $(N+1)^2$. This threadblock is then divided into several 32-lane-wide `warps' on the NVIDIA Tesla V100, and 64-lane-wide `wavefronts' on the AMD GPUs. Any additional threads in excess of the threadblock size are simply masked out when executing on the hardware. Each warp/wavefront then has access to a number of vector registers. On the NVIDIA Tesla V100, a warp is able to reserve a maximum of 255 128B vector registers (each register being  32 lanes of 4B words) within a Stream Multiprocessor (SM) on the GPU. Similarly, on the AMD GPUs each wavefront is able to reserve a maximum of 256 256B vector registers (each register being 64 lanes of 4B words) in each Compute Unit (CU). While terminology between the GPU vendors differ, they share the property that warps/wavefronts reserving large amounts of register space can substantially affect the number of warps/waves which can be simultaneously active on an SM/CU. Each GPU's SMs/CUs have vector register files that are divided evenly between four SIMD units, with both the NVIDIA Tesla V100 SMs and the AMD Instinct MI100 CUs having a total of 256KB of vector register space, and the AMD Instinct MI250X CUs having 512KB of vector register space. The `occupancy' of a kernel then refers to the average number of warps/wavefronts which are active on each SM/CU during the kernel's execution. Other constraints, such as the requirement that all warps/wavefronts from a given threadblock must occupy the same SM/CU can also restrict the occupancy of a kernel.

We list in Table \ref{tab:operator_reg} some metrics regarding the occupancy of the Poisson operator kernel on each of the GPUs considered above. For each threadblock size, we list the number of warps/wavefronts required to decompose the block. We then list the number of vector registers required by each warp/wavefront to execute the kernel, information we extract from the assembly generated by the compiler. Using these values, we list the occupancy of the operator kernel both in terms of warps/wavefronts active per SM/CU, and in terms of how many spectral elements each SM/CU processes while occupied. 

\begin{table*}
\centering
\begin{tabular}{@{}ccccccccc@{}} \toprule
\multicolumn{2}{c}{} & \multicolumn{7}{c}{$N$} \\ \cmidrule{3-9}
\multicolumn{2}{c}{} & 9 & 10 & 11 & 12 & 13 & 14 & 15 \\ \cmidrule(l){2-9}
\multicolumn{1}{c}{} & Block size, $(N+1)^2$ & 100 & 121 & 144 & 169 & 196 & 225 & 256 \\ \cmidrule{1-9}
\multirow{4}{*}{NVIDIA Tesla V100} 
& Warps/Block          &  4 &  4 &   5 &  6 &  7 &   8 &   8 \\
& Registers/Warp       & 80 & 80 & 104 & 90 & 82 & 102 & 100 \\
& Occupancy (Warps/SM) & 24 & 24 &  15 & 18 & 14 &  16 & 16 \\
& Elements/SM          &  6 &  6 &   3 &  3 &  2 &   2 &  2 \\
\midrule
\multirow{4}{*}{AMD Instinct MI100} 
& Wavefronts/Block          &   2 &   2 &   3 &   3 &   4 &   4 &   4 \\
& Registers/Wavefront       & 113 & 125 & 125 & 125 & 125 & 125 & 125 \\
& Occupancy (Wavefronts/CU) &   8 &   8 &   6 &   6 &   8 &   8 &   8 \\
& Elements/CU               &   4 &   4 &   2 &   2 &   2 &   2 &   2 \\
\midrule
\multirow{4}{*}{AMD Instinct MI250X} 
& Wavefronts/Block          &   2 &   2 &   3 &   3 &   4 &   4 &   4 \\
& Registers/Wavefront       & 126 & 124 & 124 & 124 & 124 & 124 & 124 \\
& Occupancy (Wavefronts/CU) &  16 &  16 &  12 &  12 &  16 &  16 &  16 \\
& Elements/CU               &   8 &   8 &   4 &   4 &   4 &   4 &   4 \\
\bottomrule
\end{tabular}
\caption{Occupancy metrics for 2D threadblock algorithm for the Poisson operator kernel in hipBone. The number of warps/wavefronts must be sufficiently large to cover the threadblock size. Each warp/wavefront then uses some number of vector registers, which is the primary factor limiting SM/CU occupancy. The occupancy in terms of warps/SM or wavefronts/CU is listed for each kernel, along with the occupancy in terms of elements per SM/CU.}
\label{tab:operator_reg}
\end{table*}

Comparing the occupancy metrics of the operator kernel on the AMD Instinct MI100 in Table \ref{tab:operator_reg} to the achieved performance in Figure \ref{MI100Operator.fig}, we see that the drop in performance observed at $N=11$ corresponds to a drop in occupancy on each CU. The occupancy drop itself is caused by the threadblock size of of 144 requiring one additional wavefront at the same register usage compared to $N=10$. This third wavefront will also unfortunately have 48 of its lanes masked while executing. The subsequent $N=12$ kernel on the AMD Instinct MI100 achieves the same occupancy as $N=11$, and the performance sees a increase which follows the trend of the roofline model, but shares the same inefficiencies as $N=11$, albeit with less masked lanes. At $N=13$ and onward, the kernel requires four wavefronts per threadblock and occupancy in terms of wavefronts/CU is recovered. The kernels, however, now do more computation with more wavefronts synchronizing their activity to coordinate accesses to shared memory. The performance continues to track upwards, but the inefficiency compared to the roofline model is never recovered. 

\begin{figure*}
\centering
    \begin{subfigure}[t]{0.45\textwidth}
        \centering
        \begin{tikzpicture}[scale=0.7]
\begin{axis}[
  xmode=log, 
  ymode=log, 
  grid=both, 
  major grid style={line width=.1pt,draw=gray!50}, 
  minor grid style={line width=.1pt,draw=gray!50}, 
  domain=1:8, 
  width=4.5in, 
  ymin=1e-16, 
  xlabel={Degrees of Freedom}, 
  ylabel={hipBone GFLOPS},
  cycle list name=will, 
  legend cell align=left, 
  legend pos=north west, 
  mark size=1.5pt, 
  line width=1.4pt, 
  legend entries={N1n1,N1n3,N1n6,N2n12,N4n24,N8n48},
  title={Nvidia Tesla V100, $N$=7},
  ymax=1.0e5,
  ymin=10.0,
  xmin=1e4,
  xmax=1e10,
]
\addplot   table {
39304 61.3
328509 332.7
1124864 494.2
2685619 573.4
5268024 616.2
9129329 631.6
14526784 649.3
21717639 654.5
30959144 652.6
42508549 657.6
};
\addplot   table {
120224 68.2
995049 511.7
3396224 1116.8
8095499 1473.4
15864624 1675.1
27475349 1817.0
43699424 1876.9
65308599 1892.4
93074624 1920.8
127769249 1934.3
};
\addplot   table {
243984 111.0
2004519 777.6
6825104 2093.9
16249239 2284.3
31820424 3290.7
55082159 3560.0
87577944 3682.2
130851279 3772.7
186445664 3810.5
255904599 3848.8
};
\addplot   table {
495144 206.0
4038089 1331.1
13715834 2016.3
32615379 3890.0
63823724 4999.7
110427869 6058.0
175514814 6662.2
262171559 7050.9
373485104 7386.2
512542449 7485.8
};
\addplot   table {
997464 294.8
8105229 1251.1
27497294 3503.5
65347659 6411.1
127830324 9088.5
221119289 10811.5
351388554 12398.4
524812119 13432.3
747563984 14311.6
1025818149 14648.4
};
\addplot   table {
2009384 743.8
16268769 4659.5
55126154 6959.8
130929539 11813.5
256026924 18248.1
442766309 21784.4
703495694 24252.9
1050563079 26517.2
1496316464 28310.6
2053103849 29126.9
};
\end{axis}
\end{tikzpicture}
        \caption{HipBone FOM in GLFOPS on Summit for a variety of problem sizes using polynomial degree $N=7$.}
        \label{SummitScalingP7.fig}
    \end{subfigure}
    \hfill
    \begin{subfigure}[t]{0.45\textwidth}
        \centering
        \begin{tikzpicture}[scale=0.7]
\begin{axis}[
  xmode=log, 
  grid=both, 
  major grid style={line width=.1pt,draw=gray!50}, 
  minor grid style={line width=.1pt,draw=gray!50}, 
  domain=1:8, 
  width=4.5in, 
  ymin=1e-16, 
  xlabel={Degrees of Freedom per Rank}, 
  ylabel={Degrees of Freedom * Iterations / (Ranks * Time)},
  cycle list name=will, 
  legend cell align=left, 
  legend pos=north west, 
  mark size=1.5pt, 
  line width=1.4pt, 
  legend entries={N1n1,N1n3,N1n6,N2n12,N4n24,N8n48},
  title={Nvidia Tesla V100, $N$=7},
  ymax=5.5e9,
  ymin=0.0,
  xmin=1e4,
  xmax=1e8,
]
\addplot   table {
39304.0 289000000.0
328509.0 1640000000.0
1124864.0 2470000000.0
2685619.0 2890000000.0
5268024.0 3120000000.0
9129329.0 3210000000.0
14526784.0 3310000000.0
21717639.0 3340000000.0
30959144.0 3330000000.0
42508549.0 3360000000.0
};
\addplot   table {
40074.666666666664 109000000.0
331683.0 850000000.0
1132074.6666666667 1880000000.0
2698499.6666666665 2490000000.0
5288208.0 2840000000.0
9158449.666666666 3090000000.0
14566474.666666666 3190000000.0
21769533.0 3220000000.0
31024874.666666668 3280000000.0
42589749.666666664 3300000000.0
};
\addplot   table {
40664.0 90400000.0
334086.5 650000000.0
1137517.3333333333 1770000000.0
2708206.5 1940000000.0
5303404.0 2800000000.0
9180359.833333334 3030000000.0
14596324.0 3140000000.0
21808546.5 3220000000.0
31074277.333333332 3250000000.0
42650766.5 3290000000.0
};
\addplot   table {
41262.0 85100000.0
336507.4166666667 561000000.0
1142986.1666666667 855000000.0
2717948.25 1650000000.0
5318643.666666667 2130000000.0
9202322.416666666 2590000000.0
14626234.5 2850000000.0
21847629.916666668 3010000000.0
31123758.666666668 3160000000.0
42711870.75 3200000000.0
};
\addplot   table {
41561.0 61400000.0
337717.875 264000000.0
1145720.5833333333 745000000.0
2722819.125 1370000000.0
5326263.5 1940000000.0
9213303.708333334 2310000000.0
14641189.75 2650000000.0
21867171.625 2870000000.0
31148499.333333332 3060000000.0
42742422.875 3140000000.0
};
\addplot   table {
41862.166666666664 78000000.0
338932.6875 494000000.0
1148461.5416666667 741000000.0
2727698.7291666665 1260000000.0
5333894.25 1950000000.0
9224298.104166666 2330000000.0
14656160.291666666 2590000000.0
21886730.8125 2840000000.0
31173259.666666668 3030000000.0
42772996.854166664 3120000000.0
};
\end{axis}
\end{tikzpicture}
        \caption{Throughput, in terms of DOFS*Iterations/(Ranks*Time), of hipBone on Summit over a variety of problem sizes using polynomial degree $N=7$.}
        \label{SummitThroughputP7.fig}
    \end{subfigure}
    \hfill
    \begin{subfigure}[t]{0.45\textwidth}
        \centering
        \begin{tikzpicture}[scale=0.7]
\begin{axis}[
  xmode=log, 
  ymode=log, 
  grid=both, 
  major grid style={line width=.1pt,draw=gray!50}, 
  minor grid style={line width=.1pt,draw=gray!50}, 
  domain=1:8, 
  width=4.5in, 
  ymin=1e-16, 
  xlabel={Degrees of Freedom}, 
  ylabel={hipBone GFLOPS},
  cycle list name=will, 
  legend cell align=left, 
  legend pos=north west, 
  mark size=1.5pt, 
  line width=1.4pt, 
  legend entries={N1n1,N1n3,N1n6,N2n12,N4n24,N8n48},
  title={Nvidia Tesla V100, $N$=15},
  ymax=1.0e5,
  ymin=10.0,
  xmin=1e4,
  xmax=1e10,
]
\addplot   table {
24389 40.3
205379 293.2
704969 595.6
1685159 765.3
3307949 843.9
5735339 920.5
9129329 1000.2
13651919 1030.1
19465109 1033.8
26730899 1035.9
35611289 1041.7
46268279 1051.5
};
\addplot   table {
74849 57.5
623099 439.0
2130749 1176.0
5083799 1649.3
9968249 2197.4
17270099 2517.9
27475349 2753.1
41069999 2902.3
58540049 2924.3
80371499 3010.8
107050349 3042.2
139062599 3086.1
};
\addplot   table {
152279 84.5
1256759 608.4
4285439 1969.3
10210319 2598.7
20003399 4341.9
34636679 4882.0
55082159 5503.9
82311839 5610.2
117297719 5945.5
161011799 5957.5
214426079 6079.3
278512559 6129.7
};
\addplot   table {
309809 173.7
2534819 1143.0
8619029 3072.2
20506439 3972.5
40141049 5604.9
69466859 7374.3
110427869 9167.2
164968079 10143.7
235031489 10746.5
322562099 11443.3
429503909 11709.4
557800919 11935.5
};
\addplot   table {
624869 339.1
5090939 2144.1
17286209 3270.6
41098679 7189.5
80416349 9849.1
139127219 14527.6
221119289 17372.2
330280559 19237.7
470499029 20897.3
645662699 22358.8
859659569 22990.9
1116377639 23611.4
};
\addplot   table {
1260329 624.2
10224659 3930.1
34668989 10663.0
82369319 13077.7
161101649 19089.1
278641979 27208.9
442766309 32308.7
661250639 35868.6
941870969 38964.0
1292403299 43808.3
1720623629 45294.6
2234307959 43657.7
};
\end{axis}
\end{tikzpicture}
        \caption{HipBone FOM in GLFOPS on Summit for a variety of problem sizes using polynomial degree $N=15$.}
        \label{SummitScalingP15.fig}
    \end{subfigure}
    \hfill
    \begin{subfigure}[t]{0.45\textwidth}
        \centering
        \begin{tikzpicture}[scale=0.7]
\begin{axis}[
  xmode=log, 
  grid=both, 
  major grid style={line width=.1pt,draw=gray!50}, 
  minor grid style={line width=.1pt,draw=gray!50}, 
  domain=1:8, 
  width=4.5in, 
  ymin=1e-16, 
  xlabel={Degrees of Freedom per Rank}, 
  ylabel={Degrees of Freedom * Iterations / (Ranks * Time)},
  cycle list name=will, 
  legend cell align=left, 
  legend pos=north west, 
  mark size=1.5pt, 
  line width=1.4pt, 
  legend entries={N1n1,N1n3,N1n6,N2n12,N4n24,N8n48},
  title={Nvidia Tesla V100, $N$=15},
  ymax=5.5e9,
  ymin=0.0,
  xmin=1e4,
  xmax=1e8,
]
\addplot   table {
24389.0 133000000.0
205379.0 1020000000.0
704969.0 2100000000.0
1685159.0 2720000000.0
3307949.0 3020000000.0
5735339.0 3300000000.0
9129329.0 3590000000.0
13651919.0 3710000000.0
19465109.0 3730000000.0
26730899.0 3740000000.0
35611289.0 3760000000.0
46268279.0 3800000000.0
};
\addplot   table {
24949.666666666668 64600000.0
207699.66666666666 513000000.0
710249.6666666666 1390000000.0
1694599.6666666667 1970000000.0
3322749.6666666665 2630000000.0
5756699.666666667 3020000000.0
9158449.666666666 3310000000.0
13689999.666666666 3490000000.0
19513349.666666668 3520000000.0
26790499.666666668 3630000000.0
35683449.666666664 3670000000.0
46354199.666666664 3730000000.0
};
\addplot   table {
25379.833333333332 48300000.0
209459.83333333334 358000000.0
714239.8333333334 1170000000.0
1701719.8333333333 1560000000.0
3333899.8333333335 2610000000.0
5772779.833333333 2940000000.0
9180359.833333334 3320000000.0
13718639.833333334 3380000000.0
19549619.833333332 3590000000.0
26835299.833333332 3600000000.0
35737679.833333336 3670000000.0
46418759.833333336 3710000000.0
};
\addplot   table {
25817.416666666668 50500000.0
211234.91666666666 340000000.0
718252.4166666666 920000000.0
1708869.9166666667 1190000000.0
3345087.4166666665 1690000000.0
5788904.916666667 2220000000.0
9202322.416666666 2770000000.0
13747339.916666666 3060000000.0
19585957.416666668 3250000000.0
26880174.916666668 3460000000.0
35791992.416666664 3540000000.0
46483409.916666664 3610000000.0
};
\addplot   table {
26036.208333333332 49700000.0
212122.45833333334 320000000.0
720258.7083333334 491000000.0
1712444.9583333333 1080000000.0
3350681.2083333335 1490000000.0
5796967.458333333 2190000000.0
9213303.708333334 2630000000.0
13761689.958333334 2910000000.0
19604126.208333332 3160000000.0
26902612.458333332 3380000000.0
35819148.708333336 3480000000.0
46515734.958333336 3580000000.0
};
\addplot   table {
26256.854166666668 46100000.0
213013.72916666666 294000000.0
722270.6041666666 802000000.0
1716027.4791666667 986000000.0
3356284.3541666665 1440000000.0
5805041.229166667 2060000000.0
9224298.104166666 2440000000.0
13776054.979166666 2720000000.0
19622311.854166668 2950000000.0
26925068.729166668 3320000000.0
35846325.604166664 3430000000.0
46548082.479166664 3310000000.0
};
\end{axis}
\end{tikzpicture}
        \caption{Throughput, in terms of DOFS*Iterations/(Ranks*Time), of hipBone on Summit over a variety of problem sizes using polynomial degree $N=15$.}
        \label{SummitThroughputP15.fig}
    \end{subfigure}
    \hfill
    \caption{Performance of full hipBone benchmark on the ORNL Summit cluster using NVIDIA Tesla V100 GPUs. Performance is measured over a variety of problem sizes, on 1 to 48 MPI ranks, each utilizing a single NVIDIA Tesla V100.}
    \label{fig:SummitScaling}
\end{figure*}
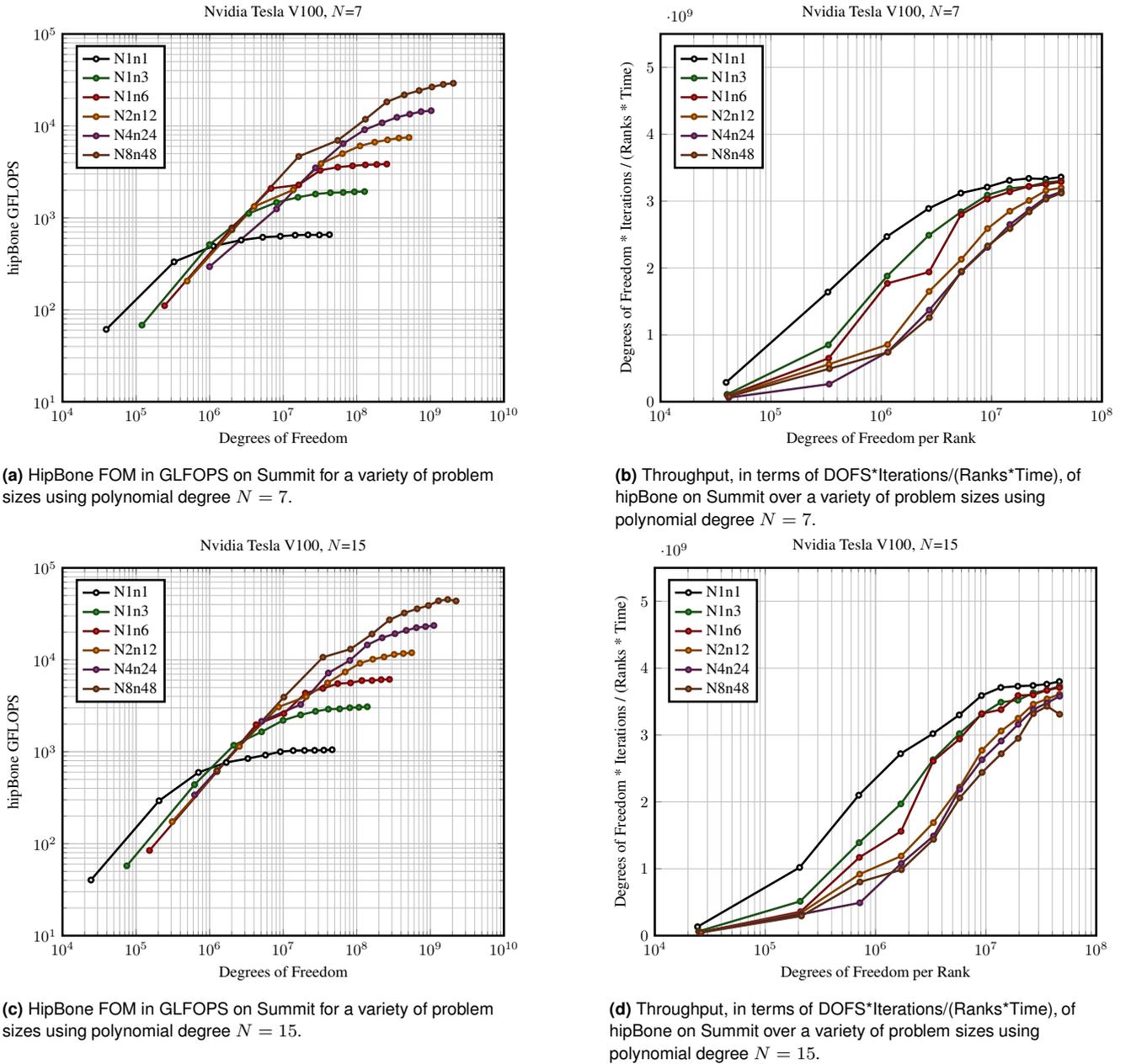

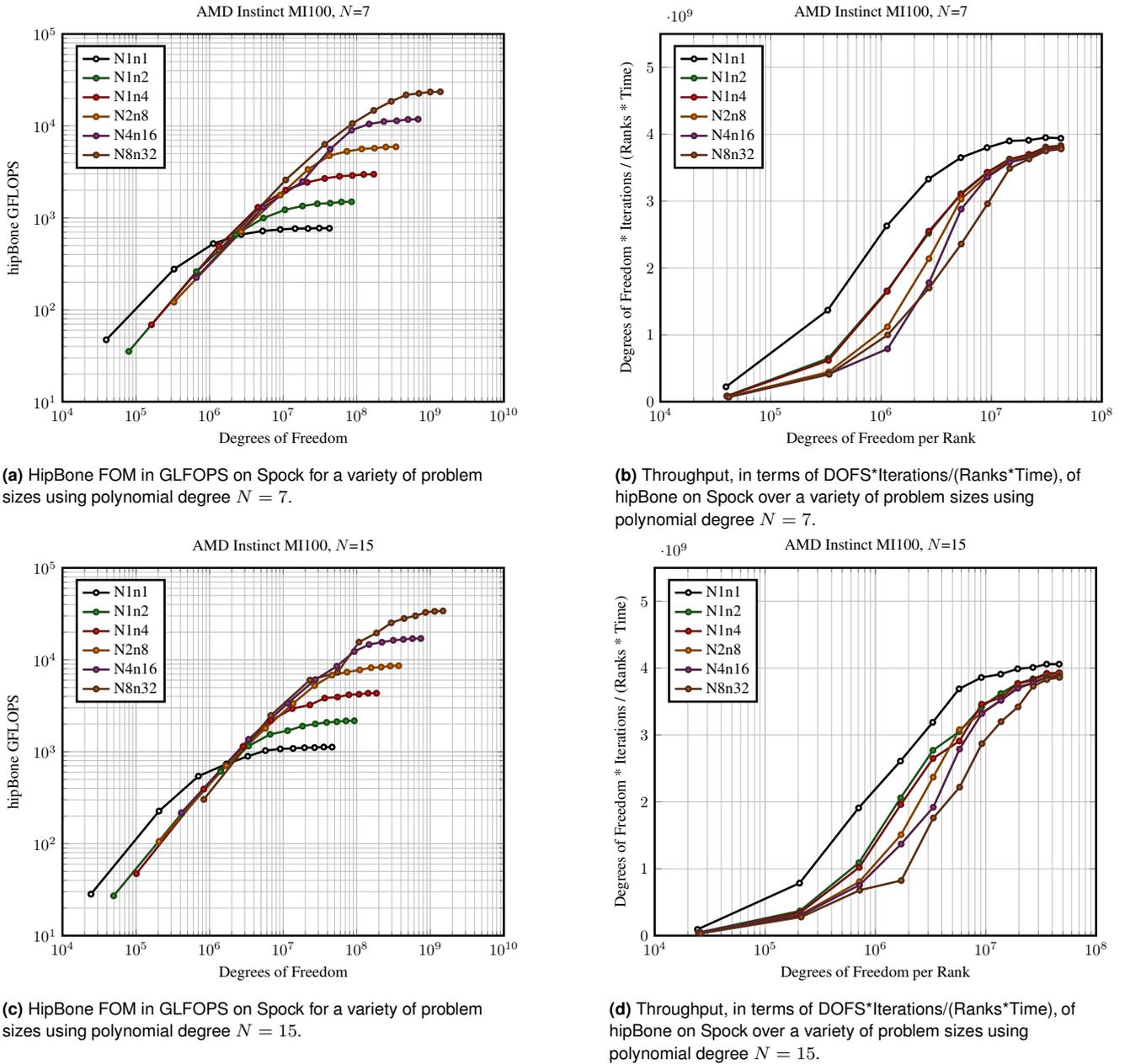
\begin{figure*}
\centering
    \begin{subfigure}[t]{0.45\textwidth}
        \centering
        \begin{tikzpicture}[scale=0.7]
\begin{axis}[
  xmode=log, 
  ymode=log, 
  grid=both, 
  major grid style={line width=.1pt,draw=gray!50}, 
  minor grid style={line width=.1pt,draw=gray!50}, 
  domain=1:8, 
  width=4.5in, 
  ymin=1e-16, 
  xlabel={Degrees of Freedom}, 
  ylabel={hipBone GFLOPS},
  cycle list name=will, 
  legend cell align=left, 
  legend pos=north west, 
  mark size=1.5pt, 
  line width=1.4pt, 
  legend entries={N1n1,N1n2,N1n4,N2n8,N4n16,N8n32},
  title={AMD Instinct MI100, $N$=7},
  ymax=1.0e5,
  ymin=10.0,
  xmin=1e4,
  xmax=1e10,
]
\addplot   table {
39304 47.2
328509 277.4
1124864 525.7
2685619 660.1
5268024 721.1
9129329 748.7
14526784 765.5
21717639 766.4
30959144 773.6
42508549 770.5
};
\addplot   table {
79764 35.3
661779 260.2
2260544 658.1
5390559 995.9
10566324 1223.7
18302339 1346.3
29113104 1423.2
43513119 1445.1
62016884 1489.6
85138899 1497.5
};
\addplot   table {
161874 68.7
1333149 493.4
4542824 1305.3
10819899 2007.7
21193374 2435.4
36692249 2685.3
58345524 2834.8
87182199 2889.9
124231274 2968.8
170521749 2982.0
};
\addplot   table {
328509 121.6
2685619 707.0
9129329 1771.2
21717639 3362.8
42508549 4751.9
73560059 5286.4
116930169 5619.0
174676879 5720.1
248858189 5893.9
341532099 5924.1
};
\addplot   table {
661779 224.8
5390559 1310.7
18302339 2485.6
43513119 5585.7
85138899 9006.0
147295679 10486.2
234099459 11165.0
349666239 11382.3
498112019 11753.3
683552799 11821.8
};
\addplot   table {
1333149 441.0
10819899 2586.5
36692249 6294.0
87182199 10619.1
170521749 14751.9
294942899 18468.2
468677649 21753.2
699957999 22597.1
997015949 23388.0
1368083499 23513.9
};
\end{axis}
\end{tikzpicture}
        \caption{HipBone FOM in GLFOPS on Spock for a variety of problem sizes using polynomial degree $N=7$.}
        \label{SpockScalingP7.fig}
    \end{subfigure}
    \hfill
    \begin{subfigure}[t]{0.45\textwidth}
        \centering
        \begin{tikzpicture}[scale=0.7]
\begin{axis}[
  xmode=log, 
  grid=both, 
  major grid style={line width=.1pt,draw=gray!50}, 
  minor grid style={line width=.1pt,draw=gray!50}, 
  domain=1:8, 
  width=4.5in, 
  ymin=1e-16, 
  xlabel={Degrees of Freedom per Rank}, 
  ylabel={Degrees of Freedom * Iterations / (Ranks * Time)},
  cycle list name=will, 
  legend cell align=left, 
  legend pos=north west, 
  mark size=1.5pt, 
  line width=1.4pt, 
  legend entries={N1n1,N1n2,N1n4,N2n8,N4n16,N8n32},
  title={AMD Instinct MI100, $N$=7},
  ymax=5.5e9,
  ymin=0.0,
  xmin=1e4,
  xmax=1e8,
]
\addplot   table {
39304.0 223000000.0
328509.0 1370000000.0
1124864.0 2630000000.0
2685619.0 3330000000.0
5268024.0 3650000000.0
9129329.0 3800000000.0
14526784.0 3900000000.0
21717639.0 3910000000.0
30959144.0 3950000000.0
42508549.0 3940000000.0
};
\addplot   table {
39882.0 84600000.0
330889.5 647000000.0
1130272.0 1660000000.0
2695279.5 2520000000.0
5283162.0 3110000000.0
9151169.5 3430000000.0
14556552.0 3630000000.0
21756559.5 3690000000.0
31008442.0 3810000000.0
42569449.5 3830000000.0
};
\addplot   table {
40468.5 83600000.0
333287.25 618000000.0
1135706.0 1650000000.0
2704974.75 2550000000.0
5298343.5 3100000000.0
9173062.25 3430000000.0
14586381.0 3620000000.0
21795549.75 3700000000.0
31057818.5 3800000000.0
42630437.25 3820000000.0
};
\addplot   table {
41063.625 75000000.0
335702.375 446000000.0
1141166.125 1120000000.0
2714704.875 2140000000.0
5313568.625 3030000000.0
9195007.375 3380000000.0
14616271.125 3600000000.0
21834609.875 3660000000.0
31107273.625 3780000000.0
42691512.375 3800000000.0
};
\addplot   table {
41361.1875 69900000.0
336909.9375 415000000.0
1143896.1875 791000000.0
2719569.9375 1780000000.0
5321181.1875 2880000000.0
9205979.9375 3360000000.0
14631216.1875 3580000000.0
21854139.9375 3650000000.0
31132001.1875 3770000000.0
42722049.9375 3790000000.0
};
\addplot   table {
41660.90625 69000000.0
338121.84375 411000000.0
1146632.78125 1000000000.0
2724443.71875 1700000000.0
5328804.65625 2360000000.0
9216965.59375 2960000000.0
14646176.53125 3490000000.0
21873687.46875 3630000000.0
31156748.40625 3750000000.0
42752609.34375 3780000000.0
};
\end{axis}
\end{tikzpicture}
        \caption{Throughput, in terms of DOFS*Iterations/(Ranks*Time), of hipBone on Spock over a variety of problem sizes using polynomial degree $N=7$.}
        \label{SpockThroughputP7.fig}
    \end{subfigure}
    \hfill
    \begin{subfigure}[t]{0.45\textwidth}
        \centering
        \begin{tikzpicture}[scale=0.7]
\begin{axis}[
  xmode=log, 
  ymode=log, 
  grid=both, 
  major grid style={line width=.1pt,draw=gray!50}, 
  minor grid style={line width=.1pt,draw=gray!50}, 
  domain=1:8, 
  width=4.5in, 
  ymin=1e-16, 
  xlabel={Degrees of Freedom}, 
  ylabel={hipBone GFLOPS},
  cycle list name=will, 
  legend cell align=left, 
  legend pos=north west, 
  mark size=1.5pt, 
  line width=1.4pt, 
  legend entries={N1n1,N1n2,N1n4,N2n8,N4n16,N8n32},
  title={AMD Instinct MI100, $N$=15},
  ymax=1.0e5,
  ymin=10.0,
  xmin=1e4,
  xmax=1e10,
]
\addplot   table {
24389 28.3
205379 226.3
704969 542.5
1685159 733.6
3307949 892.6
5735339 1030.3
9129329 1072.9
13651919 1087.1
19465109 1106.5
26730899 1110.4
35611289 1124.8
46268279 1123.4
};
\addplot   table {
49619 27.1
414239 211.8
1417859 612.5
3384479 1153.7
6638099 1547.0
11502719 1695.3
18302339 1899.2
27360959 2009.2
39002579 2085.6
53551199 2122.4
71330819 2166.0
92665439 2172.5
};
\addplot   table {
100949 47.2
835499 392.7
2851649 1144.5
6797399 2188.1
13320749 2945.8
23069699 3229.5
36692249 3832.3
54836399 3934.6
78150149 4167.3
107281499 4219.0
142878449 4329.8
185588999 4333.0
};
\addplot   table {
205379 105.9
1685159 708.0
5735339 1800.0
13651919 3351.2
26730899 5246.4
46268279 6816.5
73560059 7349.3
109902239 7762.8
156590819 8204.6
214921799 8341.5
286191179 8562.5
371694959 8609.5
};
\addplot   table {
414239 217.6
3384479 1365.3
11502719 3368.3
27360959 6083.9
53551199 8505.4
92665439 12319.0
147295679 14650.1
220033919 15542.6
313472159 16297.1
430202399 16658.6
572816639 17002.3
743906879 17065.0
};
\addplot   table {
835499 303.2
6797399 2484.4
23069699 6026.4
54836399 7289.1
107281499 15584.8
185588999 19591.8
294942899 25281.5
440527199 28172.7
627525899 30085.7
861122999 32819.3
1146502499 33695.7
1488848399 34011.7
};
\end{axis}
\end{tikzpicture}
        \caption{HipBone FOM in GLFOPS on Spock for a variety of problem sizes using polynomial degree $N=15$.}
        \label{SpockScalingP15.fig}
    \end{subfigure}
    \hfill
    \begin{subfigure}[t]{0.45\textwidth}
        \centering
        \begin{tikzpicture}[scale=0.7]
\begin{axis}[
  xmode=log, 
  grid=both, 
  major grid style={line width=.1pt,draw=gray!50}, 
  minor grid style={line width=.1pt,draw=gray!50}, 
  domain=1:8, 
  width=4.5in, 
  ymin=1e-16, 
  xlabel={Degrees of Freedom per Rank}, 
  ylabel={Degrees of Freedom * Iterations / (Ranks * Time)},
  cycle list name=will, 
  legend cell align=left, 
  legend pos=north west, 
  mark size=1.5pt, 
  line width=1.4pt, 
  legend entries={N1n1,N1n2,N1n4,N2n8,N4n16,N8n32},
  title={AMD Instinct MI100, $N$=15},
  ymax=5.5e9,
  ymin=0.0,
  xmin=1e4,
  xmax=1e8,
]
\addplot   table {
24389.0 93100000.0
205379.0 785000000.0
704969.0 1910000000.0
1685159.0 2610000000.0
3307949.0 3190000000.0
5735339.0 3690000000.0
9129329.0 3860000000.0
13651919.0 3910000000.0
19465109.0 3990000000.0
26730899.0 4010000000.0
35611289.0 4060000000.0
46268279.0 4060000000.0
};
\addplot   table {
24809.5 45400000.0
207119.5 370000000.0
708929.5 1090000000.0
1692239.5 2060000000.0
3319049.5 2770000000.0
5751359.5 3050000000.0
9151169.5 3420000000.0
13680479.5 3620000000.0
19501289.5 3770000000.0
26775599.5 3840000000.0
35665409.5 3920000000.0
46332719.5 3930000000.0
};
\addplot   table {
25237.25 40200000.0
208874.75 346000000.0
712912.25 1020000000.0
1699349.75 1960000000.0
3330187.25 2650000000.0
5767424.75 2910000000.0
9173062.25 3460000000.0
13709099.75 3560000000.0
19537537.25 3770000000.0
26820374.75 3820000000.0
35719612.25 3920000000.0
46397249.75 3930000000.0
};
\addplot   table {
25672.375 45900000.0
210644.875 315000000.0
716917.375 807000000.0
1706489.875 1510000000.0
3341362.375 2370000000.0
5783534.875 3080000000.0
9195007.375 3330000000.0
13737779.875 3520000000.0
19573852.375 3720000000.0
26865224.875 3780000000.0
35773897.375 3880000000.0
46461869.875 3910000000.0
};
\addplot   table {
25889.9375 47500000.0
211529.9375 305000000.0
718919.9375 757000000.0
1710059.9375 1370000000.0
3346949.9375 1920000000.0
5791589.9375 2790000000.0
9205979.9375 3320000000.0
13752119.9375 3520000000.0
19592009.9375 3700000000.0
26887649.9375 3780000000.0
35801039.9375 3860000000.0
46494179.9375 3880000000.0
};
\addplot   table {
26109.34375 33400000.0
212418.71875 278000000.0
720928.09375 679000000.0
1713637.46875 824000000.0
3352546.84375 1760000000.0
5799656.21875 2220000000.0
9216965.59375 2870000000.0
13766474.96875 3200000000.0
19610184.34375 3420000000.0
26910093.71875 3730000000.0
35828203.09375 3830000000.0
46526512.46875 3860000000.0
};
\end{axis}
\end{tikzpicture}
        \caption{Throughput, in terms of DOFS*Iterations/(Ranks*Time), of hipBone on Spock over a variety of problem sizes using polynomial degree $N=15$.}
        \label{SpockThroughputP15.fig}
    \end{subfigure}
    \hfill
    \caption{Performance of full hipBone benchmark on the ORNL Spock cluster using AMD Instinct MI100 GPUs. Performance is measured over a variety of problem sizes, on 1 to 32 MPI ranks, each utilizing a single AMD Instinct MI100.}
    \label{fig:SpockScaling}
\end{figure*}

The occupancy metrics of the operator kernel on the AMD Instinct MI250X in Table \ref{tab:operator_reg} follows the same progression as the AMD Instinct MI100, but overall occupancy is immediately doubled due to the 2x larger register file size per CU. We see a similar qualitative trend in the achieved performance of the operator kernel on the AMD Instinct MI250X in Figure \ref{MI250XOperator.fig} as on the AMD Instinct MI100, but the dips in performance are less pronounced and performance stays closer to the empirical roofline model for high degrees. This efficiency improvement is likely due to the CU occupancy doubling. 

Finally, comparing the performance trend of the NVIDIA Tesla V100 in Figure \ref{V100Operator.fig} to the occupancy metrics in Table \ref{tab:operator_reg}, the effects of different occupancies and registers per warp between kernels are more difficult to reason about. Overall, the NVIDIA compiler uses significantly fewer registers per warp compared with the AMD compiler, allowing for a much higher occupancy per SM, especially at the moderately high degrees of $N=9$ and 10. The occupancy decrease at $N=11$, however, appears to correspond to an increase in relative efficiency, though the occupancy increase at $N=12$ does not. The different number of registers per warp between these two kernels makes them difficult to compare, however, since the differences in the assembly generated by the compiler may have additional impacts. 

\subsection{Scaling}

Overall performance of hipBone on a single GPU device is almost completely determined by the streaming efficiency of the accelerator and the performance of the Poisson operator kernel, with latency effects such as kernel launch latency and host-to-device synchronization latency being important factors for small problem sizes. The design focus in hipBone, however, is to scale efficiently to many GPU devices by leveraging GPU-Direct RDMA technologies in GPU-Aware MPI libraries, and aggressively hiding communication times behind local computation. To demonstrate the efficacy of this design we present in this section a series of performance tests of the full hipBone benchmark, rather than simply the Poisson kernel in isolation as presented above. We scale hipBone to several MPI ranks, each using a GPU accelerator, on the ORNL Summit, Spock, and Crusher clusters, sweeping over a variety of problem sizes on several GPUs/nodes. 

\begin{figure*}
\centering
    \begin{subfigure}[t]{0.45\textwidth}
        \centering
        \begin{tikzpicture}[scale=0.7]
\begin{axis}[
  xmode=log, 
  ymode=log, 
  grid=both, 
  major grid style={line width=.1pt,draw=gray!50}, 
  minor grid style={line width=.1pt,draw=gray!50}, 
  domain=1:8, 
  width=4.5in, 
  ymin=1e-16, 
  xlabel={Degrees of Freedom}, 
  ylabel={hipBone GFLOPS},
  cycle list name=will, 
  legend cell align=left, 
  legend pos=north west, 
  mark size=1.5pt, 
  line width=1.4pt, 
  legend entries={N1n1,N1n2,N1n4,N1n8,N2n16,N4n32,N8n64},
  title={AMD Instinct MI250X, $N$=7},
  ymax=1.0e5,
  ymin=10.0,
  xmin=1e4,
  xmax=1e10,
]
\addplot   table {
39304 72.6
328509 415.3
1124864 684.9
2685619 804.4
5268024 858.7
9129329 882.6
14526784 901.2
21717639 895.4
30959144 899.2
42508549 896.5
};
\addplot   table {
79764 60.4
661779 418.0
2260544 915.6
5390559 1289.9
10566324 1495.5
18302339 1610.1
29113104 1695.5
43513119 1704.0
62016884 1739.1
85138899 1748.7
};
\addplot   table {
161874 103.2
1333149 786.6
4542824 1778.6
10819899 2500.2
21193374 2944.8
36692249 3199.2
58345524 3356.8
87182199 3387.6
124231274 3459.8
170521749 3484.0
};
\addplot   table {
328509 180.1
2685619 1356.7
9129329 3307.3
21717639 4896.7
42508549 5819.7
73560059 6344.4
116930169 6665.6
174676879 6739.1
248858189 6892.7
341532099 6932.3
};
\addplot   table {
661779 341.5
5390559 2479.2
18302339 6591.9
43513119 9500.0
85138899 11459.3
147295679 12506.2
234099459 13152.8
349666239 13339.6
498112019 13667.1
683552799 13770.2
};
\addplot   table {
1333149 646.4
10819899 4604.9
36692249 12636.0
87182199 18904.1
170521749 22714.8
294942899 24854.2
468677649 26153.6
699957999 26489.0
997015949 27063.3
1368083499 27423.2
};
\addplot   table {
2685619 1195.0
21717639 8314.9
73560059 23220.9
174676879 37175.1
341532099 45016.0
590589719 49299.4
938313739 51640.3
1401168159 52741.4
1995616979 53294.4
2738124199 54094.7
};
\end{axis}
\end{tikzpicture}
        \caption{HipBone FOM in GLFOPS on Crusher for a variety of problem sizes using polynomial degree $N=7$.}
        \label{CrusherScalingP7.fig}
    \end{subfigure}
    \hfill
    \begin{subfigure}[t]{0.45\textwidth}
        \centering
        \begin{tikzpicture}[scale=0.7]
\begin{axis}[
  xmode=log, 
  grid=both, 
  major grid style={line width=.1pt,draw=gray!50}, 
  minor grid style={line width=.1pt,draw=gray!50}, 
  domain=1:8, 
  width=4.5in, 
  ymin=1e-16, 
  xlabel={Degrees of Freedom per Rank}, 
  ylabel={Degrees of Freedom * Iterations / (Ranks * Time)},
  cycle list name=will, 
  legend cell align=left, 
  legend pos=north west, 
  mark size=1.5pt, 
  line width=1.4pt, 
  legend entries={N1n1,N1n2,N1n4,N1n8,N2n16,N4n32,N8n64},
  title={AMD Instinct MI250X, $N$=7},
  ymax=5.5e9,
  ymin=0.0,
  xmin=1e4,
  xmax=1e8,
]
\addplot   table {
39304.0 343000000.0
328509.0 2050000000.0
1124864.0 3430000000.0
2685619.0 4060000000.0
5268024.0 4350000000.0
9129329.0 4480000000.0
14526784.0 4590000000.0
21717639.0 4570000000.0
30959144.0 4590000000.0
42508549.0 4580000000.0
};
\addplot   table {
39882.0 145000000.0
330889.5 1040000000.0
1130272.0 2300000000.0
2695279.5 3260000000.0
5283162.0 3800000000.0
9151169.5 4100000000.0
14556552.0 4320000000.0
21756559.5 4350000000.0
31008442.0 4450000000.0
42569449.5 4470000000.0
};
\addplot   table {
40468.5 125000000.0
333287.25 985000000.0
1135706.0 2250000000.0
2704974.75 3180000000.0
5298343.5 3750000000.0
9173062.25 4080000000.0
14586381.0 4290000000.0
21795549.75 4330000000.0
31057818.5 4430000000.0
42630437.25 4460000000.0
};
\addplot   table {
41063.625 111000000.0
335702.375 855000000.0
1141166.125 2100000000.0
2714704.875 3120000000.0
5313568.625 3720000000.0
9195007.375 4060000000.0
14616271.125 4270000000.0
21834609.875 4320000000.0
31107273.625 4420000000.0
42691512.375 4450000000.0
};
\addplot   table {
41361.1875 106000000.0
336909.9375 784000000.0
1143896.1875 2100000000.0
2719569.9375 3030000000.0
5321181.1875 3660000000.0
9205979.9375 4000000000.0
14631216.1875 4210000000.0
21854139.9375 4280000000.0
31132001.1875 4380000000.0
42722049.9375 4420000000.0
};
\addplot   table {
41660.90625 101000000.0
338121.84375 731000000.0
1146632.78125 2020000000.0
2724443.71875 3020000000.0
5328804.65625 3640000000.0
9216965.59375 3980000000.0
14646176.53125 4190000000.0
21873687.46875 4250000000.0
31156748.40625 4340000000.0
42752609.34375 4400000000.0
};
\addplot   table {
41962.796875 94200000.0
339338.109375 662000000.0
1149375.921875 1860000000.0
2729326.234375 2980000000.0
5336439.046875 3610000000.0
9227964.359375 3960000000.0
14661152.171875 4150000000.0
21893252.484375 4240000000.0
31181515.296875 4280000000.0
42783190.609375 4350000000.0
};
\end{axis}
\end{tikzpicture}
        \caption{Throughput, in terms of DOFS*Iterations/(Ranks*Time), of hipBone on Crusher over a variety of problem sizes using polynomial degree $N=7$.}
        \label{CrusherThroughputP7.fig}
    \end{subfigure}
    \hfill
    \begin{subfigure}[t]{0.45\textwidth}
        \centering
        \begin{tikzpicture}[scale=0.7]
\begin{axis}[
  xmode=log, 
  ymode=log, 
  grid=both, 
  major grid style={line width=.1pt,draw=gray!50}, 
  minor grid style={line width=.1pt,draw=gray!50}, 
  domain=1:8, 
  width=4.5in, 
  ymin=1e-16, 
  xlabel={Degrees of Freedom}, 
  ylabel={hipBone GFLOPS},
  cycle list name=will, 
  legend cell align=left, 
  legend pos=north west, 
  mark size=1.5pt, 
  line width=1.4pt, 
  legend entries={N1n1,N1n2,N1n4,N1n8,N2n16,N4n32,N8n64},
  title={AMD Instinct MI250X, $N$=15},
  ymax=1.0e5,
  ymin=10.0,
  xmin=1e4,
  xmax=1e10,
]
\addplot   table {
24389 35.6
205379 265.3
704969 707.9
1685159 1008.0
3307949 1163.4
5735339 1289.6
9129329 1316.1
13651919 1352.9
19465109 1373.3
26730899 1377.9
35611289 1386.1
46268279 1379.6
};
\addplot   table {
49619 36.4
414239 274.2
1417859 815.3
3384479 1372.9
6638099 1817.1
11502719 2153.7
18302339 2379.2
27360959 2471.9
39002579 2594.7
53551199 2616.7
71330819 2669.6
92665439 2656.4
};
\addplot   table {
100949 69.6
835499 533.7
2851649 1576.0
6797399 2803.7
13320749 3751.3
23069699 4318.2
36692249 4722.7
54836399 4884.3
78150149 5159.1
107281499 5215.4
142878449 5316.9
185588999 5279.2
};
\addplot   table {
205379 147.2
1685159 1017.6
5735339 3022.6
13651919 5462.3
26730899 7302.6
46268279 8463.3
73560059 9395.0
109902239 9782.4
156590819 10202.5
214921799 10376.8
286191179 10540.1
371694959 10548.8
};
\addplot   table {
414239 262.6
3384479 1900.6
11502719 5600.6
27360959 10710.6
53551199 14489.2
92665439 16659.4
147295679 18397.5
220033919 19315.3
313472159 19966.6
430202399 20642.6
572816639 20913.6
743906879 20968.0
};
\addplot   table {
835499 534.7
6797399 3768.7
23069699 11120.7
54836399 20518.6
107281499 28711.2
185588999 32968.3
294942899 36574.6
440527199 38371.8
627525899 39006.9
861122999 40208.4
1146502499 41469.6
1488848399 41779.4
};
\addplot   table {
1685159 970.6
13651919 7154.0
46268279 21145.8
109902239 39986.9
214921799 56189.6
371694959 64802.6
590589719 72188.2
881974079 75644.7
1256216039 77004.2
1723683599 78534.6
2294744759 80527.2
2979767519 82375.7
};
\end{axis}
\end{tikzpicture}
        \caption{HipBone FOM in GLFOPS on Crusher for a variety of problem sizes using polynomial degree $N=15$.}
        \label{CrusherScalingP15.fig}
    \end{subfigure}
    \hfill
    \begin{subfigure}[t]{0.45\textwidth}
        \centering
        \begin{tikzpicture}[scale=0.7]
\begin{axis}[
  xmode=log, 
  grid=both, 
  major grid style={line width=.1pt,draw=gray!50}, 
  minor grid style={line width=.1pt,draw=gray!50}, 
  domain=1:8, 
  width=4.5in, 
  ymin=1e-16, 
  xlabel={Degrees of Freedom per Rank}, 
  ylabel={Degrees of Freedom * Iterations / (Ranks * Time)},
  cycle list name=will, 
  legend cell align=left, 
  legend pos=north west, 
  mark size=1.5pt, 
  line width=1.4pt, 
  legend entries={N1n1,N1n2,N1n4,N1n8,N2n16,N4n32,N8n64},
  title={AMD Instinct MI250X, $N$=15},
  ymax=5.5e9,
  ymin=0.0,
  xmin=1e4,
  xmax=1e8,
]
\addplot   table {
24389.0 117000000.0
205379.0 920000000.0
704969.0 2500000000.0
1685159.0 3580000000.0
3307949.0 4160000000.0
5735339.0 4620000000.0
9129329.0 4730000000.0
13651919.0 4870000000.0
19465109.0 4950000000.0
26730899.0 4970000000.0
35611289.0 5010000000.0
46268279.0 4990000000.0
};
\addplot   table {
24809.5 60900000.0
207119.5 479000000.0
708929.5 1450000000.0
1692239.5 2450000000.0
3319049.5 3260000000.0
5751359.5 3870000000.0
9151169.5 4290000000.0
13680479.5 4460000000.0
19501289.5 4690000000.0
26775599.5 4730000000.0
35665409.5 4830000000.0
46332719.5 4810000000.0
};
\addplot   table {
25237.25 59300000.0
208874.75 470000000.0
712912.25 1400000000.0
1699349.75 2510000000.0
3330187.25 3370000000.0
5767424.75 3890000000.0
9173062.25 4260000000.0
13709099.75 4410000000.0
19537537.25 4670000000.0
26820374.75 4720000000.0
35719612.25 4820000000.0
46397249.75 4790000000.0
};
\addplot   table {
25672.375 63800000.0
210644.875 452000000.0
716917.375 1350000000.0
1706489.875 2460000000.0
3341362.375 3290000000.0
5783534.875 3830000000.0
9195007.375 4250000000.0
13737779.875 4430000000.0
19573852.375 4620000000.0
26865224.875 4710000000.0
35773897.375 4780000000.0
46461869.875 4790000000.0
};
\addplot   table {
25889.9375 57400000.0
211529.9375 424000000.0
718919.9375 1260000000.0
1710059.9375 2420000000.0
3346949.9375 3270000000.0
5791589.9375 3770000000.0
9205979.9375 4170000000.0
13752119.9375 4380000000.0
19592009.9375 4530000000.0
26887649.9375 4680000000.0
35801039.9375 4750000000.0
46494179.9375 4760000000.0
};
\addplot   table {
26109.34375 58900000.0
212418.71875 422000000.0
720928.09375 1250000000.0
1713637.46875 2320000000.0
3352546.84375 3250000000.0
5799656.21875 3740000000.0
9216965.59375 4150000000.0
13766474.96875 4350000000.0
19610184.34375 4430000000.0
26910093.71875 4570000000.0
35828203.09375 4710000000.0
46526512.46875 4750000000.0
};
\addplot   table {
26330.609375 53900000.0
213311.234375 402000000.0
722941.859375 1190000000.0
1717222.484375 2260000000.0
3358153.109375 3180000000.0
5807733.734375 3680000000.0
9227964.359375 4100000000.0
13780844.984375 4300000000.0
19628375.609375 4370000000.0
26932556.234375 4480000000.0
35855386.859375 4580000000.0
46558867.484375 4680000000.0
};
\end{axis}
\end{tikzpicture}
        \caption{Throughput, in terms of DOFS*Iterations/(Ranks*Time), of hipBone on Crusher over a variety of problem sizes using polynomial degree $N=15$.}
        \label{CrusherThroughputP15.fig}
    \end{subfigure}
    \hfill
    \caption{Performance of full hipBone benchmark on the ORNL Crusher cluster using AMD Instinct MI250X GPUs. Performance is measured over a variety of problem sizes, on 1 to 64 MPI ranks, each utilizing a single GCD of an AMD Instinct MI250x.}
    \label{fig:CrusherScaling}
\end{figure*}
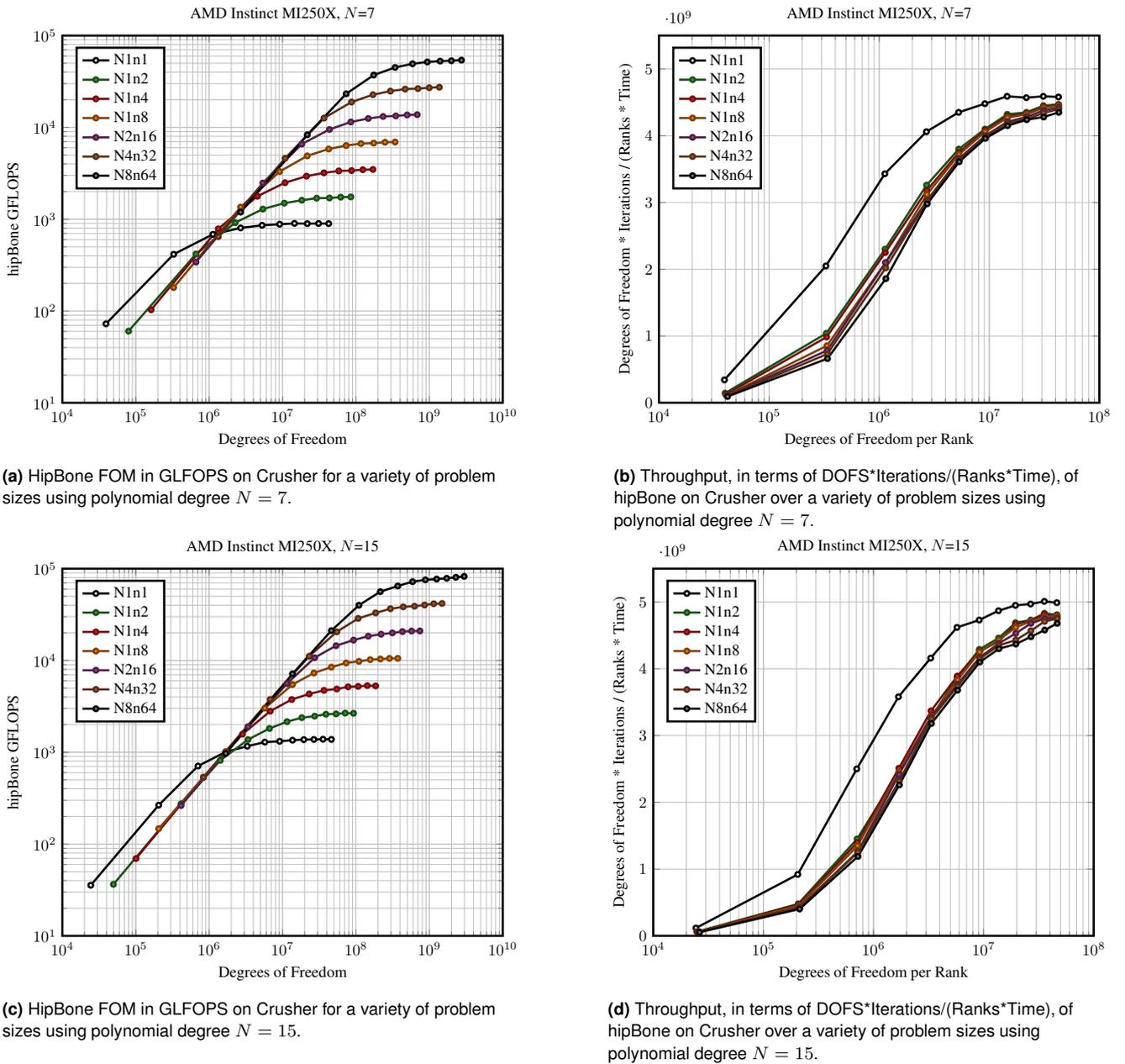

We restrict our attention to the two polynomial degrees, $N=7$ and $N=15$, for the scaling tests in this section. The choice of degree $N=15$ is of course of interest for hipBone as a benchmark, as this degree maximizes the GFLOPS of the operator kernel, and thus maximizes the overall FOM. For other polynomial degrees $N$ there will be little qualitative difference in the scaling behavior of hipBone for sufficiently large global meshes. Once there is enough local work in the Poisson operator kernels to completely hide the MPI communication time, performance (as in the single GPU case) is bound only by the asymptotic streaming rate of the GPU devices, which is not appreciably affected by $N$, and the streaming rate of the Poisson operator kernel which, as demonstrated in the tests above, is consistently near the roofline streaming rate for all $N$. For smaller problems, however, there can be differences between the scaling behaviors of different degrees due to the variety of different latency effects. We therefore include degree $N=7$ in our scaling tests, which uses the 3D threadblock algorithm for the Poisson operator kernel, so that any qualitative difference in scaling behavior can be observed.

For each test, we sweep through several global mesh sizes, ranging from the very small $\approx2500$ degrees of freedom per MPI rank, to the very large $\approx45M$ degrees of freedom per MPI rank. In this way, we collect performance data from both the extremely strong-scaled as well as weak-scaled regimes. On each cluster, we perform tests on an increasing number of GPUs, with small numbers of GPUs initially communicating entirely on-node using fast GPU-GPU data links, then over the network with GPU-Direct RDMA. Due to the different node architectures between the clusters, namely Summit nodes consisting of six GPU devices, Spock nodes consisting of four GPUs, and Crusher nodes consisting of four accelerators which appear to the OS as eight GPU devices, we scale the number of GPUs to eventually utilize a single full node, and continue scaling by using multiple full nodes. We label the configuration of each test as N$x$n$y$ to indicate the execution was on $x$ nodes of the cluster, using $y$ total GPU devices.

We show the results of the scaling tests on the ORNL Summit, Spock, and Crusher clusters in Figures \ref{fig:SummitScaling}, \ref{fig:SpockScaling}, and \ref{fig:CrusherScaling}, respectively. In each figure, we show the FOM of the hipBone application, given in GFLOPS, for $N=7$ and 15 for each problem size. As noted above, for a better comparison with historical data we use the FLOP count \eqref{eq:nekbone_flops} of the original NekBone benchmark to compute the FOM, rather than the slightly smaller true FLOP count of hipBone in \eqref{eq:hipbone_flops}.

It is difficult to immediately see how efficiently hipBone strong/weak scales from the plots of the FOM alone. One clear observation, however, is the significantly higher single rank performance of hipBone on small problem sizes. This can be attributed to some latency effects being completely avoided for single rank execution, as no MPI communication is required at all. Indeed, recalling the timeline of the Poisson operator application in Figure \ref{fig:operator_timeline}, we see that in addition to avoiding MPI latency itself, when no communication is required the two instances of host-device synchronization after packing data buffers for MPI can be skipped. Avoiding these significant sources of latency leads to better single rank performance for latency-sensitive problem sizes. 

\begin{table}
\centering
\begin{tabular}{@{}cccc@{}} \toprule
\multicolumn{1}{c}{} & Ranks & Peak FOM & FOM per Rank \\
\multicolumn{1}{c}{} &  & (GFLOPS) & (GFLOPS) \\\midrule
\multirow{6}{2cm}{Summit, NVIDIA Tesla V100}
&  1 & 1051.5 & 1051.5 \\
&  3 & 3086.1 & 1028.7 \\
&  6 & 6129.7 & 1021.6 \\
& 12 & 11935.5 & 994.6 \\
& 24 & 23611.4 & 983.8 \\
& 48 & 45294.6 & 943.6 \\
\midrule
\multirow{6}{2cm}{Spock, \;\; AMD Instinct MI100}
&  1 & 1124.8  & 1124.8 \\
&  2 & 2172.5  & 1086.3 \\
&  4 & 4333.0  & 1083.3 \\
&  8 & 8609.5  & 1076.2 \\
& 16 & 17065.0 & 1066.6 \\
& 32 & 34011.7 & 1062.9 \\
\midrule
\multirow{7}{2cm}{Crusher, AMD Instinct MI250X}
&  1 & 1386.1 & 1386.1 \\
&  2 & 2669.6 & 1334.8 \\
&  4 & 5316.9 & 1329.2 \\
&  8 & 10548.8 & 1318.6 \\
& 16 & 20968.0 & 1310.5 \\
& 32 & 41779.4 & 1305.6 \\
& 64 & 82375.7 & 1287.1 \\
\bottomrule
\end{tabular}
\caption{Summary of peak FOM in GLFOPS recorded on several multi-GPU runs of hipBone on ORNL Summit, Spock, and Crusher clusters. The far right column shows the average FOM in GLFOPS per MPI rank for each run.}
\label{table:weak_scaling_efficiency}
\end{table}

To obtain a more useful visualization of the scaling behavior of hipBone, we also show in each figure plots of the performance in terms of a `throughput` metric, similar to that used in \cite{fischer2020scalability}, defined as 
\begin{equation}\label{eq:throughput}
\mathrm{Throughput} = \frac{\mathrm{DOFs} * \mathrm{CG\;Iterations}}{\mathrm{Number\;of\;Ranks * Time}}.    
\end{equation}
This throughput metric essentially normalizes the FOM measurements by computing a rate-of-work. As the benchmark is predominately streaming bandwidth limited, and the time taken to stream data scales linearly with amount of data moved, the work done by each CG iteration therefore scales linearly with the number of degrees of freedom, $N_G$. By multiplying the global problem size with the number of CG iterations done (which is a fixed 100 iterations), the numerator of \eqref{eq:throughput} encodes a measure of the amount of work performed. The denominator of \eqref{eq:throughput} then encodes an amount of computational resources by multiplying the number of process, each with a separate GPU, by the wall-clock time required to complete the CG iterations. 

In each of the Figures \ref{fig:SummitScaling}, \ref{fig:SpockScaling}, and \ref{fig:CrusherScaling}, we show the throughput metric of each scaling test plotted against a normalized DOFs-per-rank measure. Were all tests to weak scale perfectly in all regimes, the lines on these plots would collapse to a single curve, implying the same throughput per rank can be achieved even while increasing the number of compute resources (ranks), at every problem size. This is not what is typically seen in practice, as the smaller problem sizes typically under-saturate the compute resources and communication overheads lead to lower throughput as the problem is weak-scaled. At the very large problem sizes, however, we observe a clustering of throughputs near their peak values for all numbers of ranks. This regime demonstrates the ideal weak scaling regime in which the compute resources are saturated and MPI communication is effectively hidden. 

Examining the scaling results for Summit and Spock in Figures \ref{fig:SummitScaling} and \ref{fig:SpockScaling}, respectively, we see a fair amount of variability in throughputs in the latency-sensitive strong-scaled regimes. This is somewhat expected, as we do no averaging of repeated runs on any of the clusters, and noise or jitter in communication latency, especially in multi-node tests, is normal. We see as well in the throughput plots a distinct separation of throughput curves for the multi-GPU runs which communicate entirely on-node with fast GPU-GPU links compared to the multi-node runs. Indeed, throughput on both clusters drops in the strong-scaled regime once communication travels over the network, with the throughput on Summit remaining fairly stable as the node count is increased to eight, and the throughput on the Spock cluster appearing to take additional impacts as node count increases. Comparatively, the throughput data from the Crusher cluster in Figure \ref{fig:CrusherScaling} indicates excellent scalability of hipBone to multiple MI250X GPUs. The tight clustering of all the multi-GPU throughput curves indicates very little additional latency and communication overheads as the number of ranks is increased, even to multi-node. 

In the weak scaled regime of each test, we see excellent weak scaling efficiency even compared to the single rank FOM on each GPU. We list the peak FOM of the degree $N=15$ tests in Table \ref{table:weak_scaling_efficiency} wherein we see that when weak-scaled we observe 943.6 GFLOPS or higher on each NVIDIA Tesla V100, 1062.8 GFLOPS or higher on each AMD Instinct MI100, and 1287.1 GFLOPS or higher on each GCD of AMD Instinct MI250X. Comparing to other GPU performance values for NekBone in the literature, \cite{karp2020optimization} used a version of NekBone with a native CUDA Poisson operator kernel to report $\approx410$ GFLOPS on a single NVIDIA Tesla V100 at degree $N=9$. Figure \ref{SummitScalingP7.fig} shows our hipBone benchmark exceeding this FLOP rate at the lower polynomial degree $N=7$, achieving 657.6 GFLOPS a single NVIDIA Tesla V100 despite the lower arithmetic intensity. Performance on an NVIDIA Tesla K20X was also reported by \cite{gong2016nekbone} using an OpenACC version of NekBone with a CUDA Fortran Poisson operator kernel to achieve 78.2 GFLOPS at degree $N=15$, while hipBone achieves a FLOP rate over 13 times higher the same degree on a single NVIDIA Tesla V100, despite an only 3.6x increase in peak memory bandwidth of the GPU. \cite{gong2016nekbone} also reported significant communication overhead which effected weak scaling. Using Table \ref{table:weak_scaling_efficiency}, we see that the NVIDIA Tesla V100 has weak scaling efficiency from 1 to 48 GPUs of 89.7\%, while the AMD Instinct MI100 observes a 94.5\% weak scaling efficiency from 1 to 32 GPUs, and the AMD Instinct MI250X achieves 92.9\% efficiency scaling from 1 to 64 GCDs. Considering just the scaling efficiency from two to eight nodes on each cluster, the NVIDIA Tesla V100 achieves a 94.9\% scaling efficiency, while both AMD GPUs observe greater than 98\% efficiency. This indicates that when sufficiently weak scaled, our optimizations in hipBone are very effective at hiding MPI communication overheads.

\section{Conclusion}
\label{conclusion.sec}
We have presented hipBone\footnote{\url{https://github.com/paranumal/hipbone}}, an open source and freely available GPU port of the NekBone proxy application built on top of the libParanumal finite element library and using the OCCA portability layer.  HipBone implements several optimizations over the original CPU NekBone proxy application.  Storing degrees of freedom in assembled form reduces the amount of data moved in the streaming operations as part of the CG iteration, and avoids some data motion completely.  Our GPU-aware gather-scatter library implements several nearest neighbor MPI communication algorithms and elides any explicit device-to-host data motion by leveraging GPU-Direct RDMA when possible and beneficial for performance.  HipBone also aggressively hides MPI communication in several places to improve the overall scalability of the proxy application. Firstly, the communication overhead to perform the halo exchange overlaps with the Poisson operator application kernel on one half of the interior elements. Afterwards, the gather operation needed to convert the solution vector back to global DOF layout overlaps with the Poisson operator application on the other half of the interior elements. Finally, the MPI all-reduce required for the norm calculation of the residual vector is hidden by the update of the solution vector.

In general, the hipBone proxy application code is a high-performing GPU port of the original NekBone CPU proxy application.  We have analyzed and demonstrated its performance on three systems at ORNL, each with different modern GPU accelerators.  We have presented a portable Poisson operator application kernel with near-roofline performance over a range of spectral element orders.  At low orders, performance is entirely bound by memory bandwidth.  At higher orders, performance of the Poisson operator kernel deviates slightly from the empirical roofline on some GPUs. These deviations correlate with drops in occupancy as a result of increased register pressure.

In scaling tests, we observe excellent scalability of the hipBone benchmark, especially for large problems. Scaling efficiency is especially good on the Crusher cluster, which has additional network injection bandwidth  with its four 200Gbps network cards per node. In the strong scaled regime performance becomes sensitive to latencies introduced by MPI communication overhead, GPU kernel launch, and host-device synchronization.  In the weak scaled regime, we observe the multi-node weak scaling efficiency on Summit to be 94.9\%, and more than 98\% on Spock and Crusher, up to eight compute nodes.

Currently hipBone does not replicate all the capabilities of the original NekBone benchmark. Most notably we have refrained from implementing any sort of preconditioner on the GPU, and left this topic for future study. Many optimizations we have described here, however, are easily forwarded into the libParanumal library and, by extension, NekRS. The Poisson operator kernel, gather-scatter communication library, and operator splitting algorithm can be `dropped in' to these projects with minimal effort. HipBone thus serves as an excellent proxy application for a crucial computational portion of these larger projects.

An additional future research topic is the investigation of the performance of hipBone on other vendors' hardware and/or other programming models, such as SYCL, but we note that hipBone is well-positioned to leverage such technology via the OCCA portability framework.

\section*{Acknowledgments}
This research was supported in part by the Exascale Computing Project, a collaborative effort of two U.S. Department of Energy organizations (Office of Science and the National Nuclear Security Administration) responsible for the planning and preparation of a capable exascale ecosystem, including software, applications, hardware, advanced system engineering, and early testbed platforms, in support of the nation’s exascale computing imperative.

This research was also supported in part by the John K. Costain Faculty Chair in Science at Virginia Tech. Computing facilities were furnished in part by the Advanced Research Computing group at Virginia Tech.

This research used resources of the Oak Ridge Leadership Computing Facility at the Oak Ridge National Laboratory, which is supported by the Office of Science of the U.S. Department of Energy under Contract No. DE-AC05-00OR22725.

AMD, the AMD Arrow logo, Instinct, EPYC, and combinations thereof are trademarks of Advanced Micro Devices, Inc. OpenCL is a trademark of Apple Inc. used by permission by Khronos Group, Inc. SYCL is a registered trademark of the Khronos Group, Inc. Other product names used in this publication are for identification purposes only and may be trademarks of their respective companies. 

\bibliographystyle{SageH}
\bibliography{references}

\begin{thebibliography}{40}
\providecommand{\natexlab}[1]{#1}
\providecommand{\url}[1]{\texttt{#1}}
\providecommand{\urlprefix}{URL }
\expandafter\ifx\csname urlstyle\endcsname\relax
  \providecommand{\doi}[1]{DOI:\discretionary{}{}{}#1}\else
  \providecommand{\doi}{DOI:\discretionary{}{}{}\begingroup
  \urlstyle{rm}\Url}\fi

\bibitem[{Anderson et~al.(2021)Anderson, Andrej, Barker, Bramwell, Camier,
  Cerveny, Dobrev, Dudouit, Fisher, Kolev et~al.}]{anderson2021mfem}
Anderson R, Andrej J, Barker A, Bramwell J, Camier JS, Cerveny J, Dobrev V,
  Dudouit Y, Fisher A, Kolev T et~al. (2021) {MFEM}: A modular finite element
  methods library.
\newblock \emph{Computers \& Mathematics with Applications} 81: 42--74.

\bibitem[{Beams et~al.(2020)Beams, Abdelfattah, Tomov, Dongarra, Kolev and
  Dudouit}]{beams2020high}
Beams N, Abdelfattah A, Tomov S, Dongarra J, Kolev T and Dudouit Y (2020)
  High-order finite element method using standard and device-level batch {GEMM}
  on {GPU}s.
\newblock In: \emph{2020 IEEE/ACM 11th Workshop on Latest Advances in Scalable
  Algorithms for Large-Scale Systems (ScalA)}. IEEE, pp. 53--60.

\bibitem[{Brown(2020)}]{brown2020exploring}
Brown N (2020) Exploring the acceleration of {N}ekbone on reconfigurable
  architectures.
\newblock In: \emph{2020 IEEE/ACM International Workshop on Heterogeneous
  High-performance Reconfigurable Computing (H2RC)}. IEEE, pp. 19--28.

\bibitem[{Canuto et~al.(2012)Canuto, Hussaini, Quarteroni, Thomas~Jr
  et~al.}]{canuto2012spectral}
Canuto C, Hussaini MY, Quarteroni A, Thomas~Jr A et~al. (2012) \emph{Spectral
  methods in fluid dynamics}.
\newblock Springer Science \& Business Media.

\bibitem[{Cebamanos et~al.(2014)Cebamanos, Henty, Richardson and
  Hart}]{cebamanos2014auto}
Cebamanos L, Henty D, Richardson H and Hart A (2014) Auto-tuning an {O}pen{ACC}
  accelerated version of {N}ek5000.
\newblock In: \emph{International conference on exascale applications and
  software}. Springer, pp. 69--81.

\bibitem[{Chalmers et~al.(2020)Chalmers, Karakus, Austin, Swirydowicz and
  Warburton}]{libparanumal2020}
Chalmers N, Karakus A, Austin AP, Swirydowicz K and Warburton T (2020)
  {libParanumal}: a performance portable high-order finite element library.
\newblock \doi{10.5281/zenodo.4004744}.
\newblock \urlprefix\url{https://github.com/paranumal/libparanumal}.
\newblock Release 0.4.0.

\bibitem[{Chalmers and Warburton(2020)}]{chalmers2020portable}
Chalmers N and Warburton T (2020) Portable high-order finite element kernels
  {I}: Streaming operations.
\newblock \emph{arXiv preprint arXiv:2009.10917} .

\bibitem[{Deville et~al.(2002)Deville, Fischer and Mund}]{deville2002high}
Deville MO, Fischer PF and Mund EH (2002) \emph{High-order methods for
  incompressible fluid flow}, volume~9.
\newblock Cambridge university press.

\bibitem[{Dong et~al.(2014)Dong, Dobrev, Kolev, Rieben, Tomov and
  Dongarra}]{dong2014step}
Dong T, Dobrev V, Kolev T, Rieben R, Tomov S and Dongarra J (2014) A step
  towards energy efficient computing: Redesigning a hydrodynamic application on
  {CPU}-{GPU}.
\newblock In: \emph{2014 IEEE 28th International Parallel and Distributed
  Processing Symposium}. IEEE, pp. 972--981.

\bibitem[{Fischer et~al.(2021)Fischer, Kerkemeier, Min, Lan, Phillips,
  Rathnayake, Merzari, Tomboulides, Karakus, Chalmers and Warburton}]{nekrs}
Fischer P, Kerkemeier S, Min M, Lan YH, Phillips M, Rathnayake T, Merzari E,
  Tomboulides A, Karakus A, Chalmers N and Warburton T (2021) Nek{RS}, a
  {GPU}-accelerated spectral element {N}avier-{S}tokes solver.
\newblock \emph{arXiv preprint arXiv:2104.05829} .

\bibitem[{Fischer et~al.(2008{\natexlab{a}})Fischer, Lottes and
  Kerkemeier}]{nek5000}
Fischer P, Lottes J and Kerkemeier S (2008{\natexlab{a}}) {N}ek5000 {W}eb page.
\newblock \urlprefix\url{http://nek5000.mcs.anl.gov}.

\bibitem[{Fischer et~al.(2008{\natexlab{b}})Fischer, Lottes, Pointer and
  Siegel}]{fischer2008petascale}
Fischer P, Lottes J, Pointer D and Siegel A (2008{\natexlab{b}}) Petascale
  algorithms for reactor hydrodynamics.
\newblock In: \emph{Journal of Physics: Conference Series}, volume 125, no. 1.
  IOP Publishing, p. 012076.

\bibitem[{Fischer et~al.(2020)Fischer, Min, Rathnayake, Dutta, Kolev, Dobrev,
  Camier, Kronbichler, Warburton, {\'S}wirydowicz
  et~al.}]{fischer2020scalability}
Fischer P, Min M, Rathnayake T, Dutta S, Kolev T, Dobrev V, Camier JS,
  Kronbichler M, Warburton T, {\'S}wirydowicz K et~al. (2020) Scalability of
  high-performance {PDE} solvers.
\newblock \emph{The International Journal of High Performance Computing
  Applications} 34(5): 562--586.

\bibitem[{G{\"o}ddeke et~al.(2007{\natexlab{a}})G{\"o}ddeke, Strzodka,
  Mohd-Yusof, McCormick, Buijssen, Grajewski and Turek}]{goddeke2007exploring}
G{\"o}ddeke D, Strzodka R, Mohd-Yusof J, McCormick P, Buijssen SH, Grajewski M
  and Turek S (2007{\natexlab{a}}) Exploring weak scalability for {FEM}
  calculations on a {GPU}-enhanced cluster.
\newblock \emph{Parallel Computing} 33(10-11): 685--699.

\bibitem[{G{\"o}ddeke et~al.(2007{\natexlab{b}})G{\"o}ddeke, Strzodka and
  Turek}]{goddeke2007performance}
G{\"o}ddeke D, Strzodka R and Turek S (2007{\natexlab{b}}) Performance and
  accuracy of hardware-oriented native-, emulated-and mixed-precision solvers
  in {FEM} simulations.
\newblock \emph{International Journal of Parallel, Emergent and Distributed
  Systems} 22(4): 221--256.

\bibitem[{Gong et~al.(2016)Gong, Markidis, Laure, Otten, Fischer and
  Min}]{gong2016nekbone}
Gong J, Markidis S, Laure E, Otten M, Fischer P and Min M (2016) Nekbone
  performance on {GPU}s with {O}pen{ACC} and {CUDA} {F}ortran implementations.
\newblock \emph{The Journal of Supercomputing} 72(11): 4160--4180.

\bibitem[{Gong et~al.(2014)Gong, Markidis, Schliephake, Laure, Henningson,
  Schlatter, Peplinski, Hart, Doleschal, Henty et~al.}]{gong2014nek5000}
Gong J, Markidis S, Schliephake M, Laure E, Henningson D, Schlatter P,
  Peplinski A, Hart A, Doleschal J, Henty D et~al. (2014) Nek5000 with
  {O}pen{ACC}.
\newblock In: \emph{International Conference on Exascale Applications and
  Software}. Springer, pp. 57--68.

\bibitem[{Henderson and Karniadakis(1995)}]{henderson1995unstructured}
Henderson RD and Karniadakis GE (1995) Unstructured spectral element methods
  for simulation of turbulent flows.
\newblock \emph{Journal of Computational Physics} 122(2): 191--217.

\bibitem[{Hockney(1982)}]{hockney1982characterization}
Hockney RW (1982) Characterization of parallel computers and algorithms.
\newblock \emph{Computer Physics Communications} 26(3-4): 285--291.

\bibitem[{Hockney(1985)}]{hockney1985r}
Hockney RW (1985) (r$\infty$, n12, s12) measurements on the 2-{CPU} {CRAY}
  {X-MP}.
\newblock \emph{Parallel Computing} 2(1): 1--14.

\bibitem[{Ivanov et~al.(2015)Ivanov, Gong, Akhmetova, Peng, Markidis, Laure,
  Machado, Rahn, Bartsch, Hart et~al.}]{ivanov2015evaluation}
Ivanov I, Gong J, Akhmetova D, Peng IB, Markidis S, Laure E, Machado R, Rahn M,
  Bartsch V, Hart A et~al. (2015) Evaluation of parallel communication models
  in {N}ek{B}one, a {N}ek5000 mini-application.
\newblock In: \emph{2015 IEEE International Conference on Cluster Computing}.
  IEEE, pp. 760--767.

\bibitem[{Karp et~al.(2020)Karp, Jansson, Podobas, Schlatter and
  Markidis}]{karp2020optimization}
Karp M, Jansson N, Podobas A, Schlatter P and Markidis S (2020) Optimization of
  tensor-product operations in {N}ek{B}one on {GPU}s.
\newblock \emph{arXiv preprint arXiv:2005.13425} .

\bibitem[{Kl{\"o}ckner et~al.(2009)Kl{\"o}ckner, Warburton, Bridge and
  Hesthaven}]{klockner2009nodal}
Kl{\"o}ckner A, Warburton T, Bridge J and Hesthaven JS (2009) Nodal
  discontinuous {G}alerkin methods on graphics processors.
\newblock \emph{Journal of Computational Physics} 228(21): 7863--7882.

\bibitem[{Komatitsch et~al.(2009)Komatitsch, Mich{\'e}a and
  Erlebacher}]{komatitsch2009porting}
Komatitsch D, Mich{\'e}a D and Erlebacher G (2009) Porting a high-order
  finite-element earthquake modeling application to {NVIDIA} graphics cards
  using {CUDA}.
\newblock \emph{Journal of Parallel and Distributed Computing} 69(5): 451--460.

\bibitem[{Kronbichler et~al.(2018)Kronbichler, Allalen, Ohlerich and
  Wall}]{kronbichler2018architecture}
Kronbichler M, Allalen M, Ohlerich M and Wall WA (2018) Which architecture is
  better suited for matrix-free finite-element algorithms: {I}ntel {S}kylake or
  {NVIDIA} {V}olta?
\newblock \emph{Supercomputing} .

\bibitem[{Kronbichler et~al.(2017)Kronbichler, Ljungkvist, Allalen, Ohlerich,
  Pasichnyk and Wall}]{kronbichler2017performance}
Kronbichler M, Ljungkvist K, Allalen M, Ohlerich M, Pasichnyk I and Wall WA
  (2017) Performance optimization of matrix-free finite-element algorithms
  within deal.{II}.
\newblock \emph{Supercomputing} .

\bibitem[{Lamb et~al.(1988)Lamb, Fox, Johnson, Lyzenga and
  Otto}]{lamb1988solving}
Lamb DA, Fox GC, Johnson MH, Lyzenga G and Otto S (1988) \emph{Solving Problems
  on Concurrent Processors: General techniques and regular problems}, volume~1.
\newblock Prentice Hall.

\bibitem[{Lottes and Fischer(2005)}]{lottes2005hybrid}
Lottes JW and Fischer PF (2005) Hybrid multigrid/{S}chwarz algorithms for the
  spectral element method.
\newblock \emph{Journal of Scientific Computing} 24(1): 45--78.

\bibitem[{Markidis et~al.(2015)Markidis, Gong, Schliephake, Laure, Hart, Henty,
  Heisey and Fischer}]{markidis2015openacc}
Markidis S, Gong J, Schliephake M, Laure E, Hart A, Henty D, Heisey K and
  Fischer P (2015) {O}pen{ACC} acceleration of the {N}ek5000 spectral element
  code.
\newblock \emph{The International Journal of High Performance Computing
  Applications} 29(3): 311--319.

\bibitem[{Medina et~al.(2014)Medina, St-Cyr and Warburton}]{occa}
Medina DS, St-Cyr A and Warburton T (2014) {OCCA}: A unified approach to
  multi-threading languages.

\bibitem[{NekBone()}]{nekbone}
NekBone (2018) {N}ek{B}one: {P}roxy app for {N}ek5000.
\newblock \urlprefix\url{https://github.com/Nek5000/Nekbone}.

\bibitem[{Otero et~al.(2019)Otero, Gong, Min, Fischer, Schlatter and
  Laure}]{otero2019openacc}
Otero E, Gong J, Min M, Fischer P, Schlatter P and Laure E (2019) Open{ACC}
  acceleration for the {PN}--{PN}-2 algorithm in {N}ek5000.
\newblock \emph{Journal of Parallel and Distributed Computing} 132: 69--78.

\bibitem[{Patera(1984)}]{patera1984spectral}
Patera AT (1984) A spectral element method for fluid dynamics: laminar flow in
  a channel expansion.
\newblock \emph{Journal of computational Physics} 54(3): 468--488.

\bibitem[{Remacle et~al.(2016)Remacle, Gandham and Warburton}]{remacle2016gpu}
Remacle JF, Gandham R and Warburton T (2016) {GPU} accelerated spectral finite
  elements on all-hex meshes.
\newblock \emph{Journal of Computational Physics} 324: 246--257.

\bibitem[{Schliephake and Laure(2014)}]{schliephake2014performance}
Schliephake M and Laure E (2014) Performance analysis of irregular collective
  communication with the crystal router algorithm.
\newblock In: \emph{International Conference on Exascale Applications and
  Software}. Springer, pp. 130--140.

\bibitem[{Stilwell(2013)}]{stilwell2013gnek}
Stilwell N (2013) \emph{gNek: A {GPU} accelerated incompressible Navier Stokes
  solver}.
\newblock Rice University.

\bibitem[{{\'S}wirydowicz et~al.(2019){\'S}wirydowicz, Chalmers, Karakus and
  Warburton}]{swirydowicz2019acceleration}
{\'S}wirydowicz K, Chalmers N, Karakus A and Warburton T (2019) Acceleration of
  tensor-product operations for high-order finite element methods.
\newblock \emph{The International Journal of High Performance Computing
  Applications} 33(4): 735--757.

\bibitem[{Tufo(1998)}]{tufo1998algorithms}
Tufo HM (1998) \emph{Algorithms for large-scale parallel simulation of unsteady
  incompressible flows in three-dimensional complex geometries}.
\newblock Brown University.

\bibitem[{Tufo and Fischer(1999)}]{tufo1999terascale}
Tufo HM and Fischer PF (1999) Terascale spectral element algorithms and
  implementations.
\newblock In: \emph{Proceedings of the 1999 ACM/IEEE Conference on
  Supercomputing}. pp. 68--es.

\bibitem[{Vincent et~al.(2021)Vincent, Gong, Karp, Peplinski, Jansson, Podobas,
  Jocksch, Yao, Hussain, Markidis et~al.}]{vincent2021strong}
Vincent J, Gong J, Karp M, Peplinski A, Jansson N, Podobas A, Jocksch A, Yao J,
  Hussain F, Markidis S et~al. (2021) Strong scaling of {O}pen{ACC} enabled
  {N}ek5000 on several {GPU} based {HPC} systems.
\newblock \emph{arXiv preprint arXiv:2109.03592} .

\end{thebibliography}

\end{document}